%% file: SMP-24-009_temp.tex
\pdfoutput=1
\documentclass[11pt,twoside,a4paper,cmspaper,final,collab]{cms-tdr}
\def\svnVersion{808090f}\def\svnDate{2025/07/24}\def\cmsCernNoTag{CERN-EP-2024-314}\def\cmsCernDate{\today}\def\cmsMessage{Published in Physics Letters B as \href{http://dx.doi.org/10.1016/j.physletb.2025.139754}{\doi{10.1016/j.physletb.2025.139754}.}}
\begin{document}\cmsNoteHeader{SMP-24-009}

\newlength\cmsTabSkip\setlength{\cmsTabSkip}{1ex}
\providecommand{\cmsTable}[1]{\resizebox{\textwidth}{!}{#1}}
\ifthenelse{\boolean{cms@external}}{\providecommand{\cmsTableA}[1]{#1}}{\providecommand{\cmsTableA}[1]{\resizebox{\textwidth}{!}{#1}}}
\newcommand{\pp}{\ensuremath{{\Pp\Pp}}\xspace}
\newcommand{\rts}{\ensuremath{\sqrt{s}}\xspace}
\newcommand{\MT}{\ensuremath{m_\text{T}}\xspace}
\newcommand{\Iso}{\ensuremath{I^{\ell}}\xspace}
\newcommand{\Imu}{\ensuremath{I^{\PGm}}\xspace}
\newcommand{\Mlb}{\ensuremath{m_{\ell\PQb}}\xspace}
\newcommand{\Rcw}{\ensuremath{R_\PQc^{\PW}}\xspace}
\newcommand{\Wtocq}{\ensuremath{\PW\to\PQc\PQq}\xspace}
\newcommand{\Wptocq}{\ensuremath{\PW^+\to\PQc\PAQq}\xspace}
\newcommand{\Wmtocq}{\ensuremath{\PW^-\to\PAQc\PQq}\xspace}
\newcommand{\Wtoqq}{\ensuremath{\PW\to\PQq\PAQq^{'}}\xspace}
\newcommand{\Wtouq}{\ensuremath{\PW\to\PQu\PQq}\xspace}
\newcommand{\brratio}{\ensuremath{\mathcal{B}(\Wtocq)/\mathcal{B}(\Wtoqq)}\xspace}
\newcommand{\Mtop}{\ensuremath{m_\PQt}\xspace}
\newcommand{\Mjjb}{\ensuremath{m_\text{jjb}^\PQt}\xspace}
\newcommand{\mWjj}{\ensuremath{m_\text{jj}^\PW}\xspace}
\newcommand{\Vj}{\ensuremath{\PV+\text{jets}}\xspace}
\newcommand{\gs}{\ensuremath{\PGg^*}\xspace}
\newcommand{\jet}{\text{jet}}
\DeclareRobustCommand{\PGLpmc}{{\HepParticle{\PGL}{c}{\pm}}\Xspace}
\newcommand{\bb}{\ensuremath{\PQb\PAQb}\xspace}
\newcommand{\PYTHIAeight}{\PYTHIA{8}\xspace}
\newcommand{\lj}{\ensuremath{\ell+\text{jets}}\xspace}
\newcommand{\OSSS}{\ensuremath{\text{OS}-\text{SS}}\Xspace}
\newcommand{\OSeqSS}{\ensuremath{\text{OS}=\text{SS}}\Xspace}

\cmsNoteHeader{SMP-24-009}
\title{Measurement of the \texorpdfstring{\PW}{W} boson decay branching fraction ratio \texorpdfstring{\brratio}{B(W to cq)/B(W to qq)} in proton-proton collisions at \texorpdfstring{$\sqrt{s} = 13\TeV$}{sqrt(s) = 13 TeV}}
\date{\today}

\abstract{
The most precise measurement to date of the \PW boson hadronic decay branching fraction ratio $\Rcw=\brratio$ is presented. The measurement is based on a sample of proton-proton collision data from the CERN LHC collected by the CMS experiment at a center-of-mass energy of 13\TeV in 2016--2018  with an integrated luminosity of $138\fbinv$. The large cross section of top quark-antiquark production at the LHC offers a sizable high-purity sample of \PW bosons suitable for this measurement. Events with one charged lepton (electron or muon) and at least four jets, two tagged as bottom quark jets, are analyzed. Charm jets are tagged using the presence of a muon inside the jet. The result,  $\Rcw=0.489 \pm 0.020$, is consistent with the standard model prediction and is twice as precise as the current world-average value. 
}

\hypersetup{
pdfauthor={CMS Collaboration},
pdftitle={Measurement of the decay branching fraction ratio B(W to cq)/B(W to qq) in proton-proton collisions at sqrt{s} = 13 TeV},
pdfsubject={CMS},
pdfkeywords={CMS, SMP, charm}}

\maketitle

\section{Introduction~\label{sec:intro}}

In this Letter, we present a measurement of the rate of charm (\PQc) quark production in \PW boson decays relative to the rate of \PW hadronic decays to various quark flavors, \Rcw= \brratio. Here \Wtocq stands for both \Wptocq (with \PAQq = \PAQd, \PAQs, or \PAQb) and \Wmtocq (with \PQq = \PQd, \PQs, or \PQb). The \Wtoqq hadronic decays include \Wtouq and \Wtocq contributions.

This measurement tests the universality of the weak interaction in the quark sector, a fundamental property of the standard model (SM) that assumes the sum of the couplings of any up-type quark to all down-type quarks is the same for all generations. This property is encoded in the Cabibbo--Kobayashi--Maskawa (CKM) matrix~\cite{Cabibbo:1963,Kobayashi:1973} through the unitarity condition $\sum_{j}\abs{V_{{ij}}}^{2} = 1$, separately for each generation \textit{i}.   

The \Rcw ratio can be expressed as a function of the 
CKM matrix elements through the relation
\begin{linenomath} 
    \label{eq:CKM}
    \ifthenelse{\boolean{cms@external}}
    {
    \begin{multline}
       \Rcw = \frac{\mathcal{B}(\Wtocq)}{\mathcal{B}(\Wtouq)+\mathcal{B}(\Wtocq)} = \\ \frac{\abs{V_{\PQc\PQd}}^2 + \abs{V_{\PQc\PQs}}^2 + \abs{V_{\PQc\PQb}}^2}{\abs{V_{\PQu\PQd}}^2 + \abs{V_{\PQu\PQs}}^2 + \abs{V_{\PQu\PQb}}^2 + \abs{V_{\PQc\PQd}}^2 + \abs{V_{\PQc\PQs}}^2 + \abs{V_{\PQc\PQb}}^2}.
    \end{multline}
    }{
    \begin{equation}
       \Rcw = \frac{\mathcal{B}(\Wtocq)}{\mathcal{B}(\Wtouq)+\mathcal{B}(\Wtocq)} = \frac{\abs{V_{\PQc\PQd}}^2 + \abs{V_{\PQc\PQs}}^2 + \abs{V_{\PQc\PQb}}^2}{\abs{V_{\PQu\PQd}}^2 + \abs{V_{\PQu\PQs}}^2 + \abs{V_{\PQu\PQb}}^2 + \abs{V_{\PQc\PQd}}^2 + \abs{V_{\PQc\PQs}}^2 + \abs{V_{\PQc\PQb}}^2}.
    \end{equation}    
    }
\end{linenomath}
Assuming the unitarity of the CKM matrix, \Rcw is expected to be equal to 1/2. Therefore, a measurement of \Rcw is a direct test of this assumption. Furthermore, a value of $\abs{V_{\PQc\PQs}}$ can be determined from Eq.~(\ref{eq:CKM}) using the \Rcw measurement and the world-average values of the other CKM matrix elements~\cite{PDG24}.

Previous measurements of \Rcw were performed by the LEP2 experiments~\cite{RcwALEPH, RcwOPAL} analyzing $\Pe^+\Pe^-\to\PW^+\PW^-$ events. The current world-average value~\cite{PDG24} derived from these measurements is $\Rcw=0.49\pm0.04$, corresponding to an uncertainty of 8\%. 

This Letter describes a measurement of \Rcw that is twice as precise as that value. The analysis is based on proton-proton ($\pp$) collision data from the CERN LHC at a center-of-mass energy $\rts=13\TeV$, collected by the CMS experiment in the period 2016--2018. 
The large cross section at the LHC for the production of top quark-antiquark pairs (\ttbar), each decaying into
a \PW boson and a bottom (\PQb) quark, offers a sizable high-purity sample of \PW bosons suitable for a high-precision study of their decays. 
 Jets are tagged as originating from the hadronization of \PQc quarks by the presence of a muon, which comes from the semileptonic decay of a \PQc hadron, inside the jet. The efficiency of this muon-based \PQc-tagging method, employed in previous CMS publications~\cite{EPJC78,CMS-PAPER-SMP-12-002,CMS-PAPER-SMP-18-013,CMS-PAPER-SMP-21-005}, can be precisely calibrated using data, enabling a precision measurement of \Rcw.  

\section{The CMS detector and object reconstruction~\label{sec:det}}
The central feature of the CMS apparatus is a superconducting solenoid of 6\unit{m} internal diameter, providing a magnetic field of 3.8\unit{T}. Within the solenoid volume are a silicon pixel and strip tracker, a lead tungstate crystal electromagnetic calorimeter (ECAL), and a brass and scintillator hadron calorimeter (HCAL), each composed of a barrel and two endcap sections. Forward calorimeters extend the pseudorapidity ($\eta$) coverage provided by the barrel and endcap detectors. Muons are measured in gas-ionization detectors embedded in the steel flux-return yoke outside the solenoid. A more detailed description of the CMS detector, along with a definition of the coordinate system used and the relevant kinematic variables, is reported in Ref.~\cite{CMS:2008xjf}.

Events of interest are selected using a two-tiered trigger system. The first level, composed of custom hardware processors, uses information from the calorimeters and muon detectors to select events at a rate of around 100\unit{kHz}, within a fixed latency of about 4\mus~\cite{CMS:2020cmk}. The second level, known as the high-level trigger, consists of a farm of processors running a version of the full event reconstruction software optimized for fast processing and reduces the event rate to around 1\unit{kHz} before data storage~\cite{CMS:2016ngn}.

A global particle-flow (PF) algorithm~\cite{CMS:2017yfk} reconstructs and identifies each individual particle in an event, with an optimized combination of information from the various elements of the CMS detector. Photons are identified as ECAL energy clusters not linked to the extrapolation of any charged-particle trajectory. Electrons are identified as a charged-particle track with many ECAL energy clusters, corresponding to the extrapolation of this track to the ECAL and possible bremsstrahlung photons emitted along the path through the tracker material. Muons are identified as tracks in the central tracker consistent with either a track or several hits in the muon system, and associated with calorimeter deposits compatible with the muon hypothesis. Charged hadrons are identified as charged-particle tracks identified neither as electrons nor as muons. Finally, neutral hadrons are identified as HCAL energy clusters not linked to any charged-hadron trajectory, or as a combined ECAL and HCAL energy excess relative to the expected charged-hadron energy deposit. 

The energy of photons is obtained from the ECAL measurement. The energy of electrons is obtained from a combination of the electron momentum at the primary vertex (PV) derived using the tracker, the energy of the corresponding ECAL cluster, and the energy sum of all bremsstrahlung photons spatially compatible with originating from the electron track. The energy resolution of reconstructed photons ranges from 1\% in the barrel section of the ECAL to approximately 3\% in the endcaps~\cite{CMS:2020uim}. For electrons with transverse momentum ($\pt$) around $45\GeV$ from $\PZ \to \Pe \Pe$ decays,  the $\pt$ resolution varies between 1.6\% and 5\%, with generally better performance in the barrel region than in the endcaps~\cite{CMS:2020uim}.
The energy of muons is obtained from the curvature of the corresponding track. Matching muons to tracks measured in the silicon tracker results in a \pt resolution of 1\% in the barrel and 3\% in the endcaps for muons with \pt up to 100\GeV~\cite{CMS:2018rym}. The energy of charged hadrons is determined from a combination of their momentum measured in the tracker and the matching calorimeter energy deposits, corrected for the response function of the calorimeters to hadronic showers. Finally, the energy of neutral hadrons is obtained from the corresponding corrected calorimeter energies. The PV is taken to be the vertex corresponding to the hardest scattering in the event, as described in Section 9.4.1 of Ref.~\cite{CMS-TDR-15-02}.  

Hadronic jets are clustered from the PF objects using the infrared- and collinear-safe anti-\kt algorithm~\cite{Cacciari:2008gp, Cacciari:2011ma} with a distance parameter of 0.4. The jet momentum is determined as the vector sum of all particle momenta in the jet, and is found from simulation to be, on average, within 5--10\% of the true momentum over the entire transverse momentum spectrum and detector acceptance. 
Additional $\pp$ interactions within the same or nearby bunch crossings (pileup) can contribute additional tracks and calorimetric energy deposits, increasing the apparent jet momentum. To mitigate this effect, charged particles identified as originating from pileup vertices are discarded, and an offset correction is applied to account for remaining contributions from neutral hadrons~\cite{CMS-PU}. Jet energy corrections are derived from simulation to bring the measured response of jets to that of particle level jets on average. In situ measurements of the momentum balance in dijet, $\text{photon} + \jet$, $\PZ + \jet$, and multijet events are used to correct for any residual differences in the jet energy scale (JES) between data and simulation~\cite{CMS:2016lmd}. The jet energy resolution (JER) typically amounts to 15--20\% at 30\GeV and 10\% at 100\GeV. A spreading procedure is applied to match the JER in simulation to that in data~\cite{CMS:2016lmd}. Additional selection criteria are applied to each jet to remove jets potentially dominated by anomalous contributions from various subdetector components or reconstruction failures.

The missing transverse momentum vector $\ptvecmiss$ is computed as the negative vector sum of the \pt of all the PF objects from the PV, including unclustered energy from the PF objects not associated with any reconstructed lepton, photon, or jet~\cite{CMS:2019ctu}. Its magnitude is denoted as $\ptmiss$, and it is a measure of the transverse momentum of particles leaving the detector undetected. The $\ptvecmiss$ is modified to correct the energy scale of the reconstructed jets in the event. Anomalous high-$\ptmiss$ events can be due to a variety of reconstruction failures, detector malfunctions, or noncollision backgrounds. Such events are rejected by event filters that are designed to identify more than 85--90\% of the spurious high-$\ptmiss$ events with a mistagging rate of less than 0.1\%~\cite{CMS:2019ctu}.

The trigger, reconstruction, and selection efficiencies are corrected in simulation to match those observed in the data. Lepton efficiencies ($\epsilon_{\ell}$) are evaluated using data samples of dilepton events in the $\PZ$ boson mass peak region with the tag-and-probe method~\cite{CMS-PAPER-EWK-10-005}, and correction factors $\epsilon_{\ell}^\text{data}/\epsilon_{\ell}^{\text{MC}}$, binned in $\pt^\ell$ and $\eta^\ell$ of the leptons, are implemented.

\section{Data and simulated event samples~\label{sec:samples}}
This measurement is performed using a data sample of $\pp$ collisions at $\rts=13\TeV$ collected by the CMS experiment during the 2016 ($36.3\fbinv$), 2017 ($41.5\fbinv$), and 2018 ($59.8\fbinv$) data-taking periods with a total integrated luminosity of approximately $138\fbinv$.

The signal process, referred to as semileptonic \ttbar, occurs in events where one of the \PW bosons produced in the decay of the \ttbar pair, decays leptonically, whereas the second \PW boson decays hadronically into two jets. The high transverse momentum lepton (electron or muon) is used for the online selection of signal events, and the identification of a \PQc jet enables the measurement of \Rcw. The presence of two additional \PQb jets from the decay of the \ttbar pair efficiently suppresses the background from events where a vector boson (\PW, \PZ/\gs) is produced in association with jets (collectively denoted as \Vj). Other background processes are \ttbar events where the two \PW bosons decay leptonically (referred to as dileptonic \ttbar) and single top quark production. The background of diboson ($\PW\PW$, $\PW\PZ$, and $\PZ\PZ$, collectively denoted as $\PV\PV$) events is negligible. 

The Monte Carlo (MC) event generator \POWHEG~v2.0~\cite{Nason:2004rx,Frixione:2007vw,Alioli:2010xd} is used to simulate the production of \ttbar and single top quark ($s$-, $t$-, and $\PQt\PW$ channels) events at next-to-leading-order (NLO) accuracy in quantum chromodynamics (QCD). The renormalization and factorization scales are set to the transverse mass $m_\text{T}^\PQt = \sqrt{\smash[b]{\Mtop^2 + \pt^2}}$ of the top quark, where $\Mtop=172.5\GeV$ is used. Simulated samples of \Vj are generated at NLO accuracy with the \MGvATNLO~\cite{Alwall} (version 2.6.3) matrix element generator. The diboson production process is modeled at leading order with samples of events generated with \PYTHIAeight~\cite{Sjostrand:2014zea} (version 8.219).
 The parton distribution functions (PDFs) NNPDF3.1 NNLO~\cite{NNPDF31nlo} are used. The output of the event generators is combined with the parton shower (PS) and hadronization simulation of \PYTHIAeight using the underlying event tune CP5~\cite{CP5}.  Pileup collisions are overlaid in each simulated event, and the generated distribution of the number of events per bunch crossing is matched to that observed in data. The detector response is simulated using \GEANTfour{}~\cite{Agostinelli:2002hh}. 

\section{Event selection~\label{sec:selection}}
The event sample is collected with a trigger algorithm~\cite{CMS:2016ngn} that requires the presence of an electron (muon) candidate with a minimum $\pt$ of 27, 32, and 32\GeV (24, 27, and 24\GeV) during the 2016, 2017, and 2018 data-taking periods, respectively. Electrons and muons are selected using tight identification criteria following the reconstruction algorithms discussed in Refs.~\cite{CMS:2020uim,CMS:2018rym}. We select electrons (muons) with $\pt^{\ell} > 35~(30)\GeV$ and $\abs{\eta^{\ell}} < 2.4$. The isolation \Iso of a lepton is defined as the scalar $\pt$ sum of all PF particles within a distance $\Delta R = \sqrt{\smash[b]{(\Delta\eta)^2 +(\Delta\phi)^2}}<0.3$ (0.4) around the electron (muon).  Electrons are identified using gradient-boosted decision trees, where isolation is one of the included observables. The muon candidate is isolated if $\Iso/\pt^{\ell} < 0.15$. Events with an additional isolated lepton with $\pt^\ell>20\GeV$ are rejected.  

The transverse mass ($\MT$) of the lepton and $\ptvecmiss$ is defined as
\begin{linenomath*}
\begin{equation*}
  \MT \equiv \sqrt{2~\pt^\ell~\ptmiss~[1-\cos(\phi_\ell-\phi_{\ptmiss})]},
\end{equation*}
\end{linenomath*}
where $\phi_\ell$ and $\phi_{\ptmiss}$ are the azimuthal angles of the lepton and the $\ptvecmiss$ vector, respectively. 
Events with $\MT < 20\GeV$ or $\ptmiss<20\GeV$ are discarded to suppress the contamination from events composed uniquely of jets produced through the strong interaction.

In addition to the requirements above that are designed to select events with a \PW boson decaying leptonically with an electron or a muon in the final state, we require the presence of at least four jets with $\pt^{\jet}>25\GeV$ and $\abs{\eta^{\jet}}<2.4$. Jets with a low angular separation between the jet axis and the selected isolated lepton, $\Delta R ({\jet},\ell)<0.4$, are not considered. 

{\tolerance=9000 
Two of the selected jets must be identified as originating from the hadronization of a \PQb quark. The \PQb tagging algorithm employs a deep neural network multiclassification algorithm (\textsc{DeepJet}~\cite{CMS-PAPER-BTV-16-002}) that considers the full information of all jet constituents, charged and neutral particles, secondary vertices, and global event variables, simultaneously~\cite{Bols:2020bkb}. The chosen working point of the algorithm corresponds to a misidentification rate of 1\% for light-quark and gluon jets (evaluated together and referred to as ``light jets'' in the following), and an identification efficiency of around 70\% for bottom quark jets. 
\par}

To further reduce the background and to improve the efficiency in the identification and classification of the four jets coming from the decay of the \ttbar pair, we impose additional kinematic constraints. 
We calculate the invariant mass of the prompt lepton and either of the two \PQb-tagged jets (\Mlb) and discard events if $\Mlb>150\GeV$ for both \PQb jet combinations. In addition, we use the \PW boson and top quark mass constraints to perform a $\chi^2$ test evaluating the compatibility of three-jet combinations with the expectation from simulation for jets originating from the decay of a top quark. The \PW boson invariant mass (\mWjj) is determined from the two highest \pt jets, excluding the \PQb jet candidates, and the top quark mass (\Mjjb) is calculated using those two jets along with one of the \PQb-tagged jets. The expected average values for \mWjj and \Mjjb, and its covariance matrix, are computed with the semileptonic \ttbar simulation. 
An event is discarded if the two possible three-jet combinations yield a p-value in the $\chi^2$ test smaller than 0.2. We also require that the \PQb jet passing the $\chi^2$ test is different than the one fulfilling the $\Mlb$ requirement. According to the simulation, after applying these kinematic requirements, the two \PQb jets are correctly matched in 98\% of cases, and the two jets from the \PW boson decay are correctly assigned in 81\%. 

The data sample selected with the above criteria, referred to as \lj, consists of 916\,680 events. According to the simulation, the data sample is composed of 45\% semileptonic \ttbar events with \Wtocq, 45\% semileptonic \ttbar with \Wtouq, 6\% dileptonic \ttbar, 3\% single top quark, and 1\% \Vj.

\subsection{Identification of charm jets~\label{sec:charm}}
The accurate identification of \PQc jets from the decay of \PW bosons, the precise calibration of the charm tagging efficiency, and the estimation of the associated systematic uncertainties are key ingredients of the \Rcw measurement. In a sizable fraction of the charm hadron decays (${\approx}$9\%~\cite{PDG24}) there is a muon in the final state. The presence of the muon inside the jet provides a high-purity signature to identify \PQc jets. Semileptonic \PQc quark decays into electrons are not considered because of the high background in identifying electrons inside jets. The muon-based charm tagging method provides a clean selection of \PQc jets with a low misidentification rate for light jets, which is precisely determined from data, as described below. This allows a precision measurement of \Rcw with reduced systematic uncertainties. 

Charm jets are tagged requiring a reconstructed muon among the constituents of either of the two selected jets associated with the \PW boson candidate decaying hadronically. The muon must satisfy the same tight reconstruction and identification quality criteria as those imposed on the muons from the \PW boson decaying leptonically, except for isolation, which is inverted requiring a minimum nonisolation of $\Imu>2.5\GeV$. The muon must be reconstructed in the region $\abs{\eta^{\PGm}} < 2.4$, with $5<\pt^{\PGm}<25\GeV$, and $\pt^{\PGm}/\pt^{\jet}<0.5$. The upper $\pt$ requirement and the $\pt^{\PGm}/\pt^{\jet}$ condition significantly reduce the contamination from prompt muons overlapping with or misreconstructed as jets. The lower $\pt$ threshold ensures a sufficiently high muon reconstruction efficiency (70\% at $\pt^{\PGm}=5\GeV$) since the muon must traverse the material in front of the muon detector and penetrate deep enough into the muon system to be reconstructed and satisfy the identification criteria. If more than one such muon is identified, the one with the highest $\pt$ is selected. 

In the semileptonic \ttbar charm signal events, the electric charges of the isolated lepton and the muon from the \PQc hadron decay are opposite. We refer to these events as opposite-sign (OS) events. We impose the OS condition as an additional selection requirement for the \PQc-tagged data sample. 

In most of the backgrounds, the number of OS events is the same as the number of events for which the charges of the isolated lepton and the muon in the jet are the same, which we refer to as same-sign (SS) events. This applies to events \PQc-tagged by the presence of a muon from the decay of a pion or a kaon within a jet, or to events \PQc-tagged by muons from \PQb jets in \ttbar events, where the production of a \bb pair from the \ttbar decay ensures symmetric contributions from OS and SS events.
Additionally, heavy quark-antiquark pairs, generated, for example, by gluon splitting in the PS and involving muons in their decay chain, also produce charge-symmetric events. 
These charge-symmetric backgrounds can be estimated from the SS data sample leveraging the \OSeqSS background symmetry.
For this reason, in the statistical analysis described in Section~\ref{sec:res}, the expectation for OS events in data will be modeled by the sum of the \OSSS subtracted simulation yields and the SS data events. This approach removes the need for the simulation to estimate the charge-symmetric background contribution. The SS data and MC yields differ by approximately 10\%.

\begin{figure*}[htbp!]
    \centering
    \includegraphics[width=0.49\textwidth]{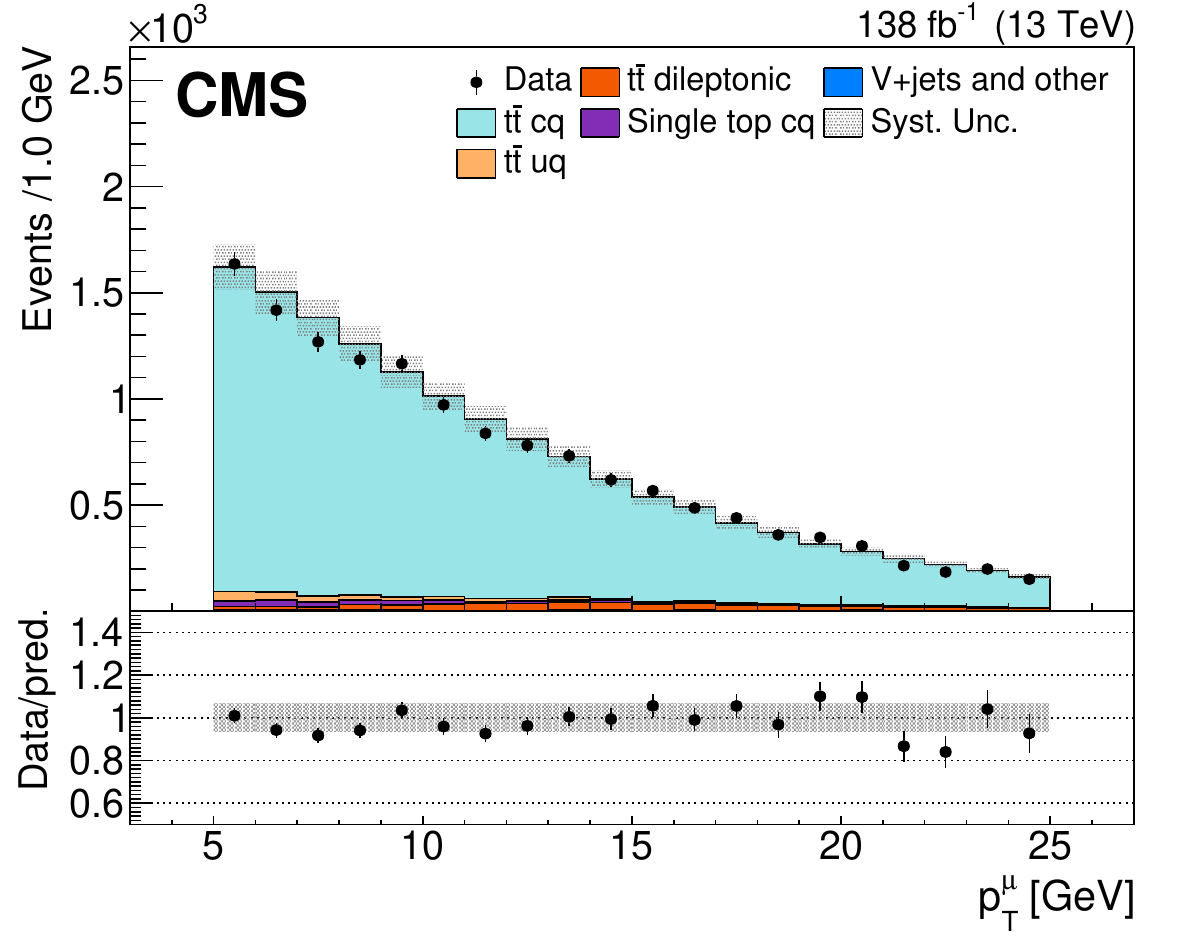}
    \includegraphics[width=0.49\textwidth]{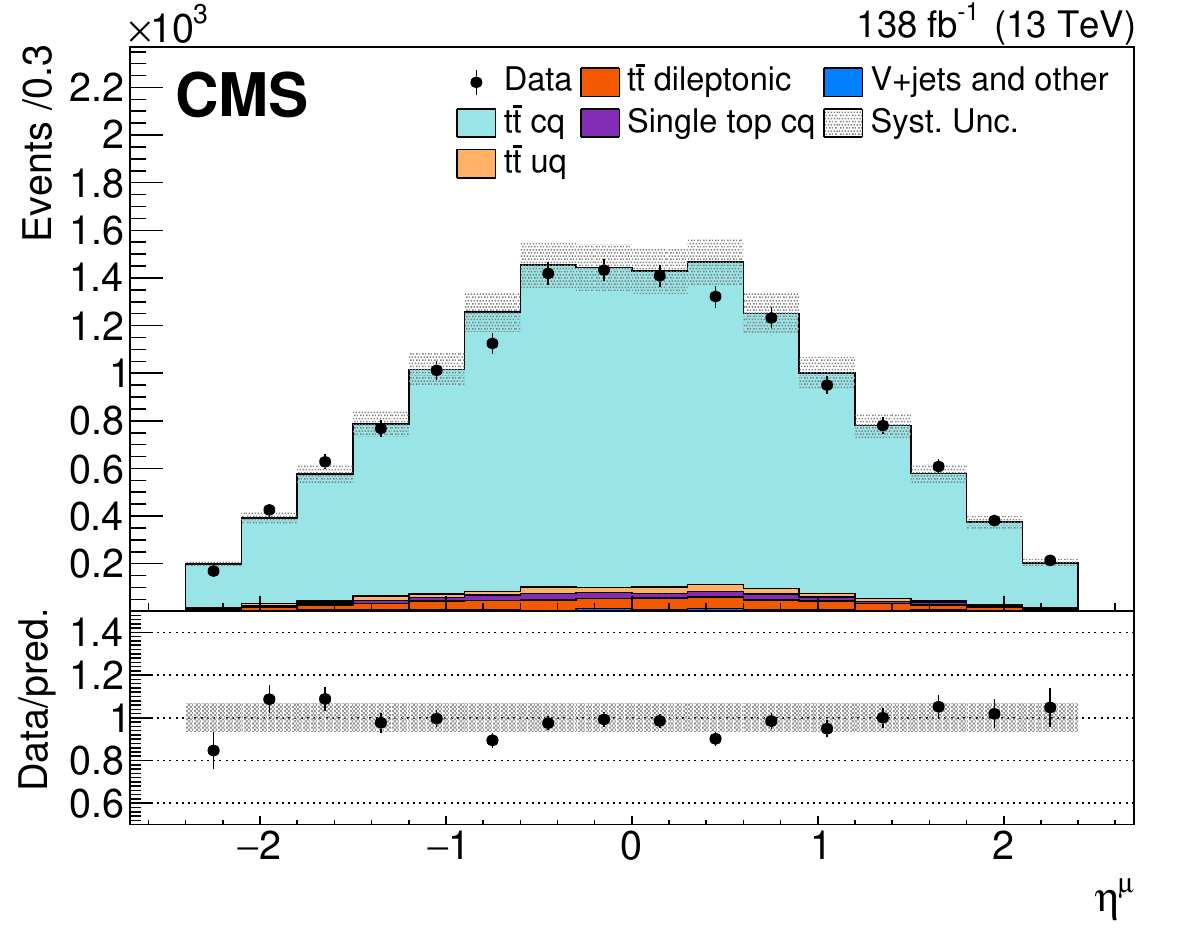}
    \includegraphics[width=0.49\textwidth]{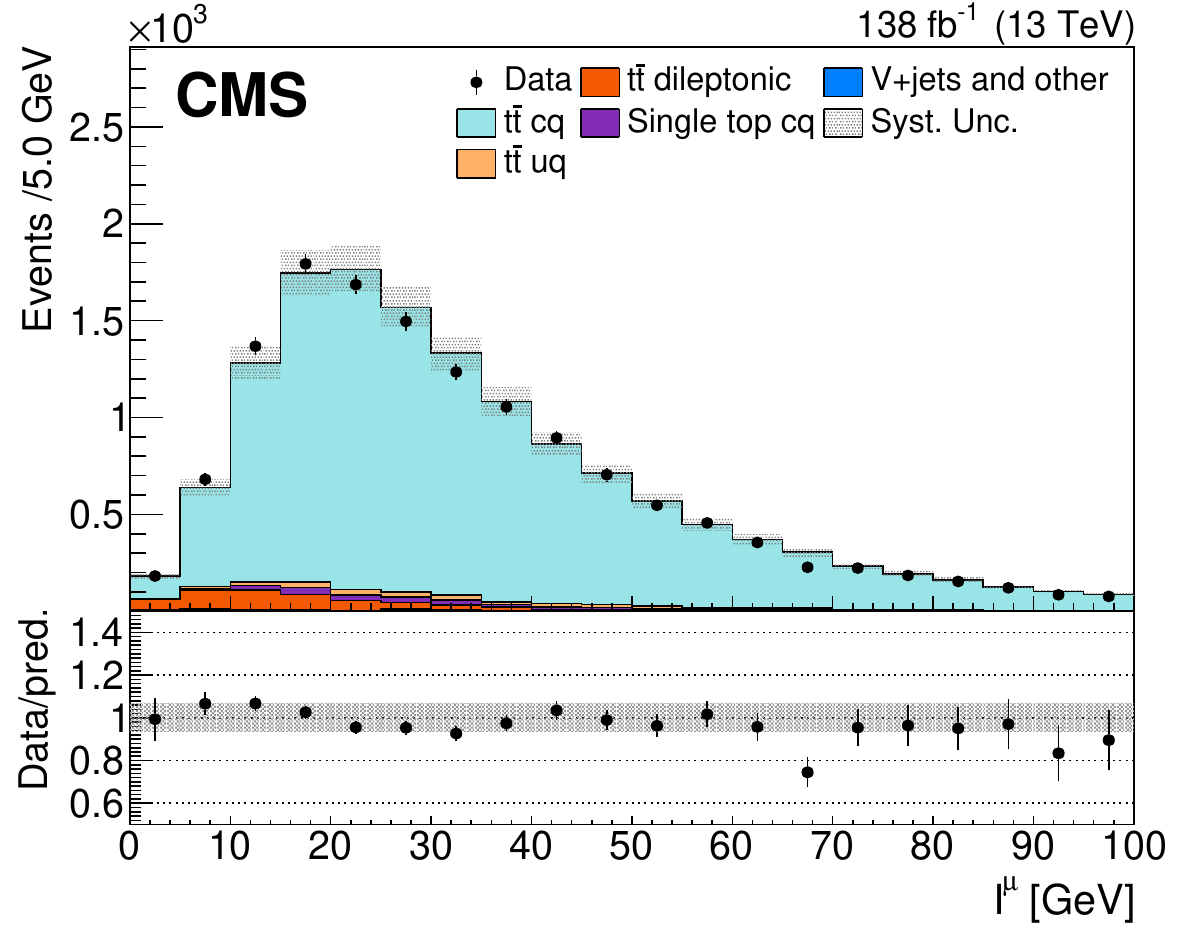}
    \includegraphics[width=0.49\textwidth]{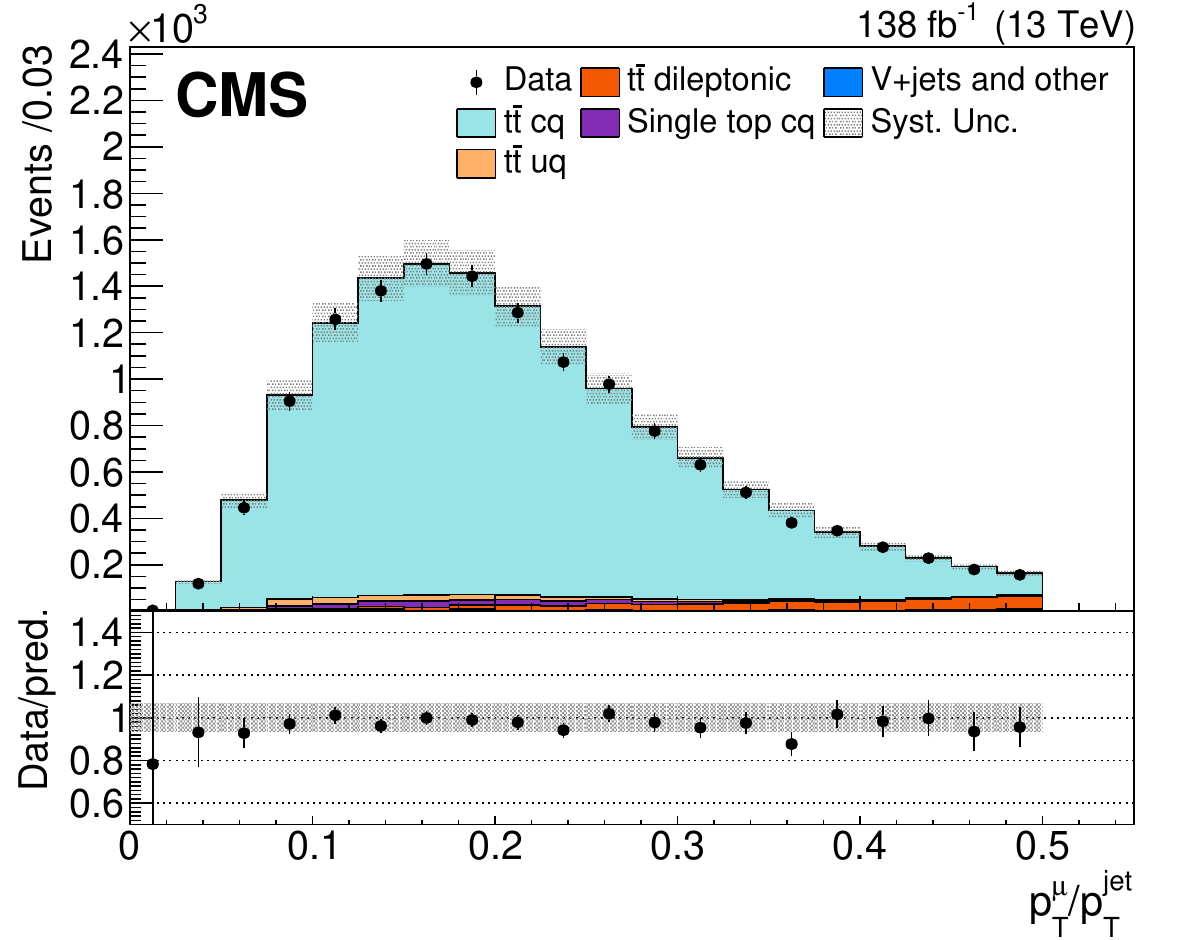}
    \caption{Kinematic distributions of the muon inside the \PQc-tagged jets: transverse momentum $\pt^{\mu}$ (upper left), pseudorapidity $\eta^{\mu}$ (upper right), isolation \Imu (lower left), and $\pt^{\mu}/\pt^{\jet}$ (lower right). Histograms show \OSSS yields for both data and simulations. Events with a prompt electron or muon are considered. The data are displayed as points with statistical error bars. The gray band in the predictions represents the systematic uncertainties. The ratios of data to the expected yields are given at the bottom of each panel, together with the statistical and systematic uncertainties.\label{fig:cmuons}}
\end{figure*}

The selected data sample containing one \PQc-tagged jet consists of 17\,973 OS and 4097 SS events. Figure~\ref{fig:cmuons} shows relevant kinematic distributions of the muon inside the tagged \PQc jets, illustrating the good agreement between data and simulation. In this figure, the normalization of the semileptonic \ttbar simulation, after applying the \lj selection criteria and before charm tagging, has been scaled to match the data.
The displayed distributions correspond to \OSSS subtracted yields, both for the data and the simulations. The gray band in the predictions accounts for the systematic uncertainties discussed in Section~\ref{sec:syst}. The \PQc-tagged jet sample has a high purity of 97\% in \PQc jets. The only significant residual contribution after \OSSS subtraction arises from dileptonic \ttbar events, where a muon from the decay of one \PW boson overlaps with a light jet. This contribution is inherently OS, since the muon inside the jet carries a charge opposite to that of the isolated lepton from the decay of the other \PW boson.

\subsubsection{Calibration of muons inside charm jets~\label{sec:muoncal}}
The modeling of the muon-based charm tagging relies on the accurate simulation of the production of \PQc hadrons from the hadronization of \PQc quarks, and their leptonic and semileptonic decays with a muon in the final state. The QCD PS, hadronization, and particle decay in the simulated samples are performed by \PYTHIAeight. The charm fragmentation fractions (FFs), defined as the probabilities for \PQc quarks to hadronize as particular \PQc hadrons, corresponding to $\PDpm$, $\PDz/\PADz$, \PDpms, and \PGLpmc hadrons, are reweighted in the MC samples to match the measured values reported in Ref.~\cite{Lisovyi:2015uqa}. These values, derived from a combination of $\Pep\Pem$, $\Pe\Pp$, and $\Pp\Pp$ collision data, are well suited to the charm production mechanism and kinematic regime relevant to this study. In addition, the leptonic and semileptonic branching fractions (BFs) of the \PQc hadrons are adjusted to agree with more recent measurements~\cite{PDG24}. The overall correction reduces the rate of muons originating from \PQc hadron decays in the simulation by 4.5\%.

The reconstruction and identification efficiency of muons inside jets in the simulation is calibrated with data. The \PQb jets in the \lj selected sample, identified with very high purity, provide a suitable source of muons for calibration. These muons, originating from the semileptonic decays of \PQb hadrons or \PQc hadrons produced in the hadronic decay of \PQb hadrons, are kinematically similar to those found in \PQc jets. The rate of muons in \PQb jets depends on several factors: the relative production rate of \PQb hadrons from \PQb quarks (i.e., the \PQb-quark FFs), the muonic BFs of \PQb hadrons, the BFs of \PQb hadrons decaying into \PQc hadrons, and the muonic BFs of the resulting \PQc hadrons. After applying the aforementioned corrections to the muonic BFs of \PQc hadrons, the muon rate in \PQb jets in the simulation aligns with experimental measurements~\cite{PDG24}, and no additional corrections are required. A correction factor for the identification efficiency of muons within jets is determined by comparing the rate of reconstructed muons inside \PQb jets in data and simulation, using the \lj selected sample as a starting point. The correction factor, which ranges from 1.0 to 0.85, is independent of $p^{\mu}_T$ but varies with muon isolation \Imu and pseudorapidity $\abs{\eta^{\PGm}}$. The correction deviates from 1 as the muon becomes less isolated and also with increasing $\abs{\eta^{\PGm}}$. The dependence on \Imu ensures that the correction derived from muons in \PQb jets remains applicable to muons in \PQc jets, which typically have lower particle multiplicity. The distributions shown in Fig.~\ref{fig:cmuons} are generated using calibrated muon efficiencies. 

\section{Systematic uncertainties~\label{sec:syst}}

Several sources of systematic uncertainty affect the predicted event yields of the various signal and background processes used to extract the \Rcw measurement. For all the uncertainties, those originating from the same source are considered fully correlated between channels, and those arising from different sources are considered uncorrelated. 

The high-\pt prompt muons and electrons in the simulated samples have uncertainties associated with the high-level trigger, reconstruction, and identification efficiencies (1--2\%). These uncertainties are uncorrelated across lepton flavors but correlated across data-taking years and are parameterized as functions of the lepton \pt and $\eta$. The uncertainty associated with the possible misidentification of the sign of the lepton electric charge is negligible. 

JES and JER uncertainties are accounted for by varying the energy of reconstructed jets in simulated events. These variations incorporate multiple uncertainty sources, which are categorized by detector regions and data-taking years~\cite{CMS:2016lmd}. The JES variations are also propagated to the \ptmiss. The resulting uncertainties in the predicted yields are 4\% (1\%) for JES (JER). The \PQb tagging identification and mistagging efficiencies in the simulation are calibrated to match the corresponding efficiencies in data~\cite{CMS-DP-2018-058}. Separate uncertainties in the tagging and mistagging corrections are assigned resulting in an overall yield uncertainty from \PQb tagging of 2.5\%. 

All MC samples are reweighted to match the pileup distribution in data, assuming a total inelastic cross section of $69.2\unit{mb}$, with a  4.6\% uncertainty~\cite{CMS:2018mlc}.  The statistical uncertainty of the charm-tagged SS data sample, which is used as the prediction for the mistag contamination of the OS signal sample, is taken as the systematic uncertainty (1.6\%) of this component. 
Additionally, the measured integrated luminosity values for the three data-taking years have uncertainties between 1.2\% and 2.5\%~\cite{CMS-LUM-17-003, CMS-PAS-LUM-17-004, CMS-PAS-LUM-18-002}, while the overall uncertainty for the combined data set is 1.6\%, affecting the predictions from the simulations.

In addition to the experimental sources, we consider theoretical uncertainties affecting the MC simulations. 
All MC samples are normalized according to their respective SM cross section values, and uncertainties in the rates of each process are assigned using the precision of CMS cross section measurements or theoretical calculations.  A 5\% uncertainty is assigned to \ttbar production~\cite{Czakon:2011xx}; 1\%, 2\%, and 4\% to single top quark $s$-, $t$-, and $\PQt\PW$-channel production~\cite{Campbell:2020fhf,Kidonakis:2021vob}; 2\% to \Vj~\cite{CMS-W-Z}; and 6\%, 5\%, and 7\% to diboson $\PW\PW$, $\PZ\PZ$, and $\PW\PZ$~\cite{CMS-PAPER-SMP-18-004,CMS-PAPER-SMP-19-001,CMS-PAPER-SMP-20-014}. 
The uncertainty from the choice of PDF is estimated from the Hessian NNPDF3.1 sets according to the procedure described in Ref.~\cite{NNPDF31nlo}; the resulting uncertainty is negligible. 
For the PS simulation, uncertainties are separately assessed for initial- and final-state radiation (ISR and FSR), by varying the respective scales up and down by factors of two. The effect of the ISR uncertainty is negligible but the FSR uncertainty changes the predicted yields of the samples with and without \PQc tagging by 6\% and 4\%, respectively, producing a significant impact on the \Rcw uncertainty of 1.9\%.
An additional uncertainty is included, which is estimated by reweighting the simulation to the measured top quark \pt in Ref.~\cite{PhysRevD.95.092001}. The effect of this uncertainty is negligible. 

The systematic effects discussed so far (except for the effect of the PS FSR uncertainty) have a minimal impact on the uncertainty of the \Rcw measurement since they affect the predictions for the \Wtocq and \Wtouq contributions in approximately the same way. The dominant source of uncertainty in \Rcw arises from systematic effects related to \PQc tagging, which influence the predicted yields of the \PQc-tagged sample.
The correction of the muon rate in the decay of \PQc hadrons in the simulation, described in Section~\ref{sec:muoncal}, introduces an uncertainty of 2.2\% in the \Wtocq rate, reflecting the precision of the charm FF and semileptonic decay BF measurements. The scale factor correcting the reconstruction and identification efficiency of muons inside jets in the simulation, determined by comparing the rate of muons in \PQb jets in data and simulation, as detailed in Section~\ref{sec:muoncal}, brings in an uncertainty of 2.7\%. This value mostly arises from the uncertainty in the measurements of the \PQb quark FFs to $\PBpm$, $\PBz/\PABz$, \PBzs, and \PGLzb hadrons, the semileptonic decay BFs of these \PQb hadrons, the decay BFs of  \PQb hadrons into $\PDpm$, $\PDz/\PADz$, \PDpms, and \PGLpmc hadrons, and the semileptonic decay BFs of these \PQc hadrons. It also includes uncertainties related to the calculation of the correction, such as residual kinematic differences between muons in \PQb jets and \PQc jets. The effect of modeling the charm quark fragmentation in the simulation is evaluated by varying $\pm 5\%$ the \pt of the muon in the \PQc-tagged jets, resulting in a systematic yield uncertainty of 0.4\%. The variation considered for $\pt^{\PGm}$ corresponds to the dispersion and uncertainty in the measurements at LEP of the average energy fraction of \PD mesons in charm quark fragmentation~\cite{charmDELPHI,charmOPAL,charmALEPH}.   

The main systematic uncertainties affecting the \Rcw measurement are summarized in Table~\ref{syst_uncert}. Relative changes in the predicted event yields of the samples with and without charm tagging are given together with the impact of each uncertainty source on the measured value of \Rcw. The total systematic uncertainty is 3.9\%.  

\begin{table*}[htbp]
 \centering
  \topcaption{Summary of the main systematic uncertainties affecting the \Rcw measurement. The quoted numbers are the percentage change in the predicted yields of the samples with no charm and with a charm tag, and the last column reflects the impacts in percentage of the measured \Rcw value from each uncertainty source. The total systematic uncertainty, calculated from the various sources of uncertainty and their correlations, is given in the last row.}
  \label{syst_uncert} 
  \renewcommand{\arraystretch}{1.2}
 \begin{tabular}{lccc}
    & No charm tag &  Charm tag &  Impact on \Rcw  \\ 
  \hline
  Source & \multicolumn{3}{c}{Uncertainty [\%]} \\
  \hline
  Charm tagging: muon identification & \NA & 2.7 & 2.6 \\
  Charm tagging: muon rate in simulation & \NA & 2.2 & 2.1 \\
  Parton shower final state radiation & 4.0 & 6.0 & 1.9 \\
  Jet energy scale & 4.0 & 4.0 & 0.6 \\
  SS data statistical uncertainty & \NA & 1.6 &  0.5\\
  Charm fragmentation modeling & \NA & 0.4 &  0.3\\
  Jet energy resolution & 1.0 & 1.0 &  0.3 \\
  \PQb tagging & 2.5 & 2.5 & 0.2 \\
  MC background normalization  & 5.0 & 5.0 & 0.1 \\
  Integrated luminosity & 1.6 & 1.6 & 0.1 \\[\cmsTabSkip]
   Total &  &  & 3.9 \\
   \end{tabular}
\end{table*}

\section{Results~\label{sec:res}}
For the extraction of the \Rcw BF ratio, the observed and predicted global event yields, split into four exclusive categories, are used to make a fit with the CMS statistical analysis and combination tool, \textsc{Combine}~\cite{combine}. The four event categories are constructed depending on the flavor of the prompt lepton (electron or muon) and the charm tagging (tag or no tag) of one of the jets associated with the \PW boson. The predicted input yields from the simulation are separated between \Wtocq and \Wtouq contributions (from both semileptonic \ttbar and single top quark production with a hadronically decaying \PW boson) and are varied in an anticorrelated manner in the fit. The fit also determines the global normalization of the combined $\Wtocq+\Wtouq$ contributions. The small background contributions from dileptonic \ttbar and \Vj events are taken from the simulation. The effects of systematic uncertainties and their correlations are included by incorporating nuisance parameters into the fit model. As noted in Section~\ref{sec:syst}, many of the systematic effects influence the normalization of the \Wtouq and \Wtocq contributions in the same way and therefore do not impact the uncertainty of the measured \Rcw ratio. 

The observed data yields entering the fit for the two categories with a charm tag are OS events. The predicted yields are constructed using the \OSSS subtracted yields from the MC simulations and the observed SS data representing the remaining background contributions, as described in Section~\ref{sec:charm}. This procedure has the advantage that the prediction for the background contribution to \Wtocq relies only on the simulation for a small fraction of the events (3\%). The remaining backgrounds are accurately characterized by data, significantly reducing the systematic uncertainty in their prediction. Table~\ref{tab:yields} lists the observed and predicted event yields for the four categories used in the fit. The uncertainties are primarily driven by correlated systematic effects, which cancel out in the determination of \Rcw.

\begin{table*}[h]
\centering
\topcaption{Observed and predicted event yields input to the fit for the four categories. Predictions are separated by process. For the two categories with charm tag, the yields predicted by the simulations correspond to \OSSS subtracted events, the SS contamination is estimated with data, and the number of observed events in data corresponds to OS events. The relative uncertainties shown in parenthesis for the predictions are based on the statistical uncertainties of the MC samples and the systematic uncertainties and their correlations discussed in Section~\ref{sec:syst}. Correlated systematic effects dominate the uncertainties.}\label{tab:yields}
\newcolumntype{x}{D{(}{\,\,(}{-1}}
\cmsTableA{
\begin{tabular}{lxxxx}
  & \multicolumn{1}{c}{Prompt \PGm}    & \multicolumn{1}{c}{Prompt \Pe}   & \multicolumn{1}{c}{Prompt \PGm}   & \multicolumn{1}{c}{Prompt \Pe}  \\
\multicolumn{1}{l}{Process}  & \multicolumn{1}{c}{no charm tag} & \multicolumn{1}{c}{no charm tag} & \multicolumn{1}{c}{charm tag} & \multicolumn{1}{c}{charm tag}    \\ \hline
\ttbar, \Wtocq         &   245\,816(7\%)  &  151\,570(7\%)  &  8172(9\%)  &  4993(9\%)   \\ 
\ttbar, \Wtouq         &   257\,789(7\%)  &  159\,146(7\%)  &  150(9\%)  &  84(9\%)     \\ 
Dileptonic \ttbar      &   31\,343(7\%)  &  19\,219(7\%)  &  299(8\%)  &  188(8\%)  \\ 
Single top, \Wtocq     &   5060(7\%)  &  3085(7\%)  &  133(10\%)  &  93(10\%)  \\ 
Single top, \Wtouq     &   4772(7\%)  &  2948(7\%)  &  2(50\%)  &  2(50\%)  \\ 
Single top, no \Wtoqq  &   3620(13\%)  &  1884(13\%)  &  15(20\%)  &  9(50\%)  \\
\Vj                    &   5005(12\%)  &  3687(12\%)  &  43(30\%)  &  9(50\%)\\ 
Diboson                &   299(12\%)  &  142(12\%)  &  1(50\%)  &  1(50\%) \\ 
[\cmsTabSkip]
Total predictions \OSSS &                  &                   &  8815(9\%)  &  5379(9\%) \\
Data SS                &                  &                   &    2551(2\%)  &  1546(2\%)   \\ 
[\cmsTabSkip]
Total predictions      &   553\,705(7\%)  &  341\,681(7\%)  &  11\,366(7\%)  &  6925(7\%) \\ 
Data OS                &  553\,378        & 341\,232          & 11\,167     & 6806      \\ 
\end{tabular}
}
\end{table*}

\begin{figure}[h]
    \centering
    \includegraphics[width=0.47\textwidth]{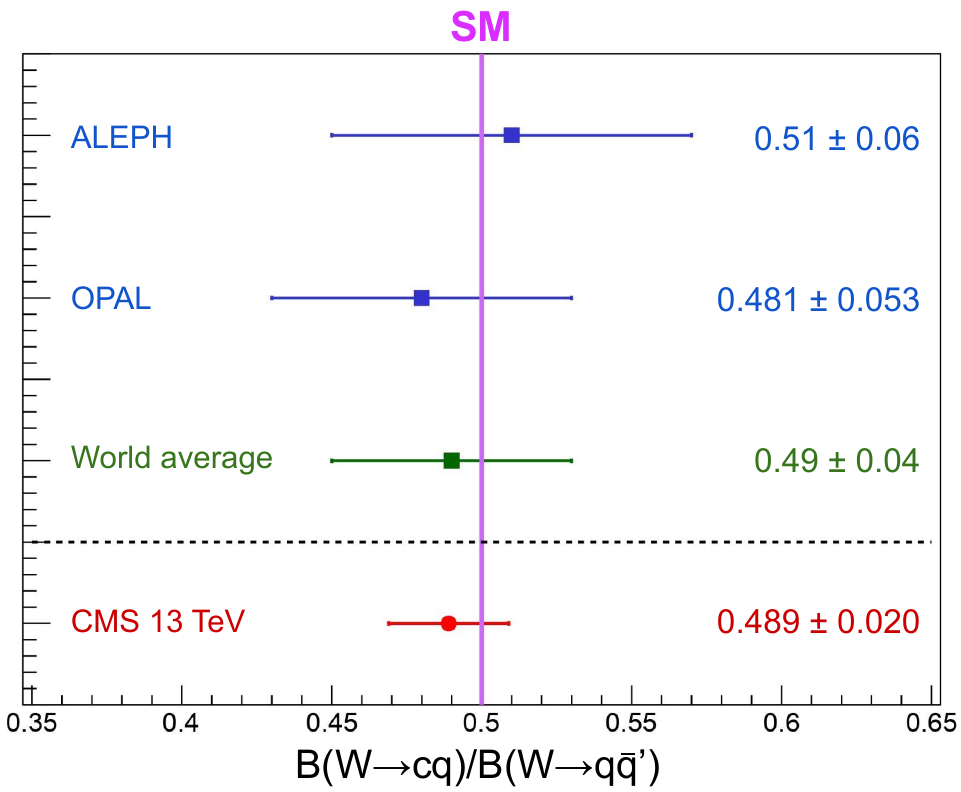}
    \caption{Comparison of the measured value of \Rcw with previous LEP2 measurements~\cite{RcwALEPH, RcwOPAL}, and the world-average value~\cite{PDG24}. Horizontal bars represent the total uncertainty of the measurements.\label{fig:Rcw}}
\end{figure}

The fit yields a value $\Rcw=0.489 \pm 0.005\stat \pm 0.019 \syst$, with a total uncertainty of $\pm 0.020$, in good agreement with the prediction of the SM. The precision of the measurement, limited by the systematic uncertainty in the charm tagging efficiency, is improved by more than a factor of two compared with the world-average value, as shown in Fig.~\ref{fig:Rcw}.  

\begin{figure*}[h!]
    \centering
    \includegraphics[width=0.49\textwidth]{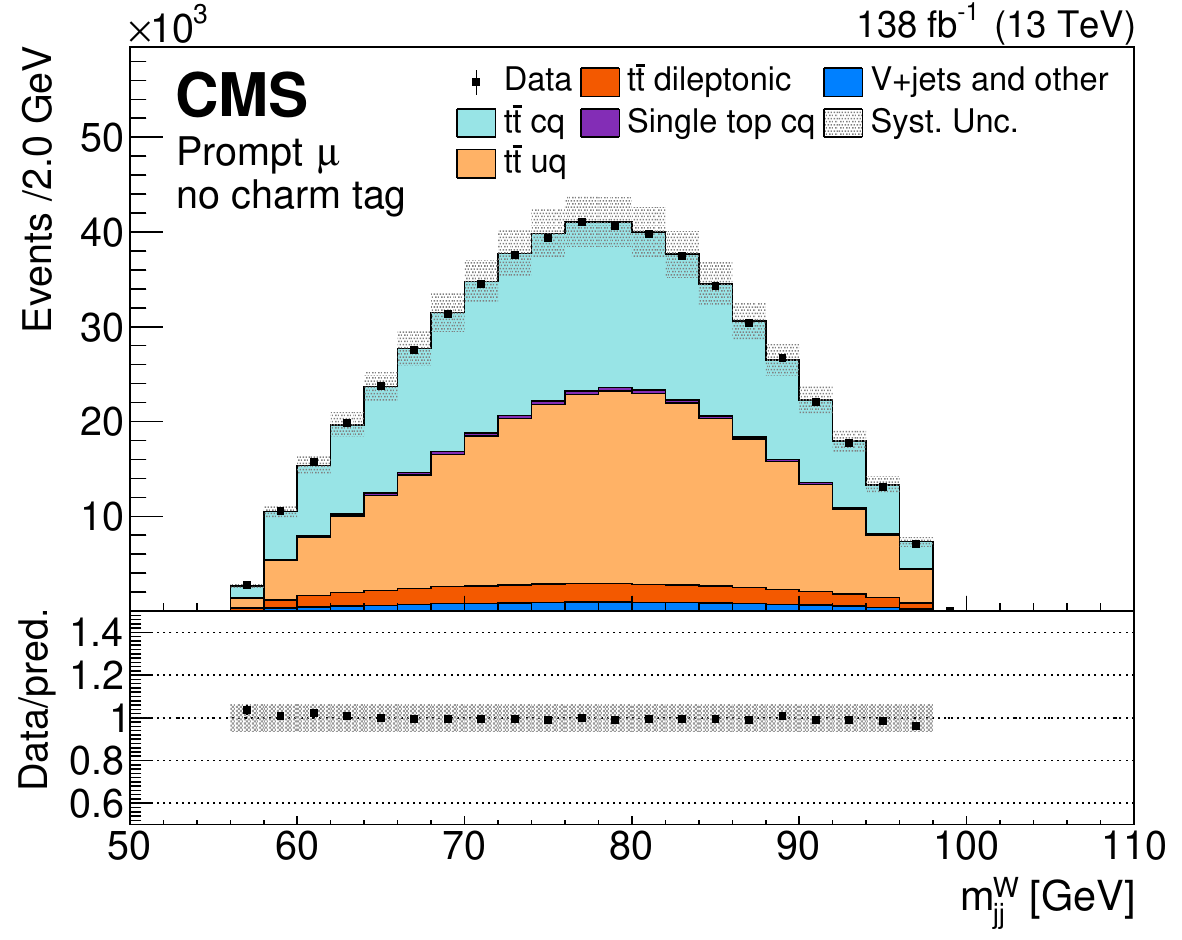}
    \includegraphics[width=0.49\textwidth]{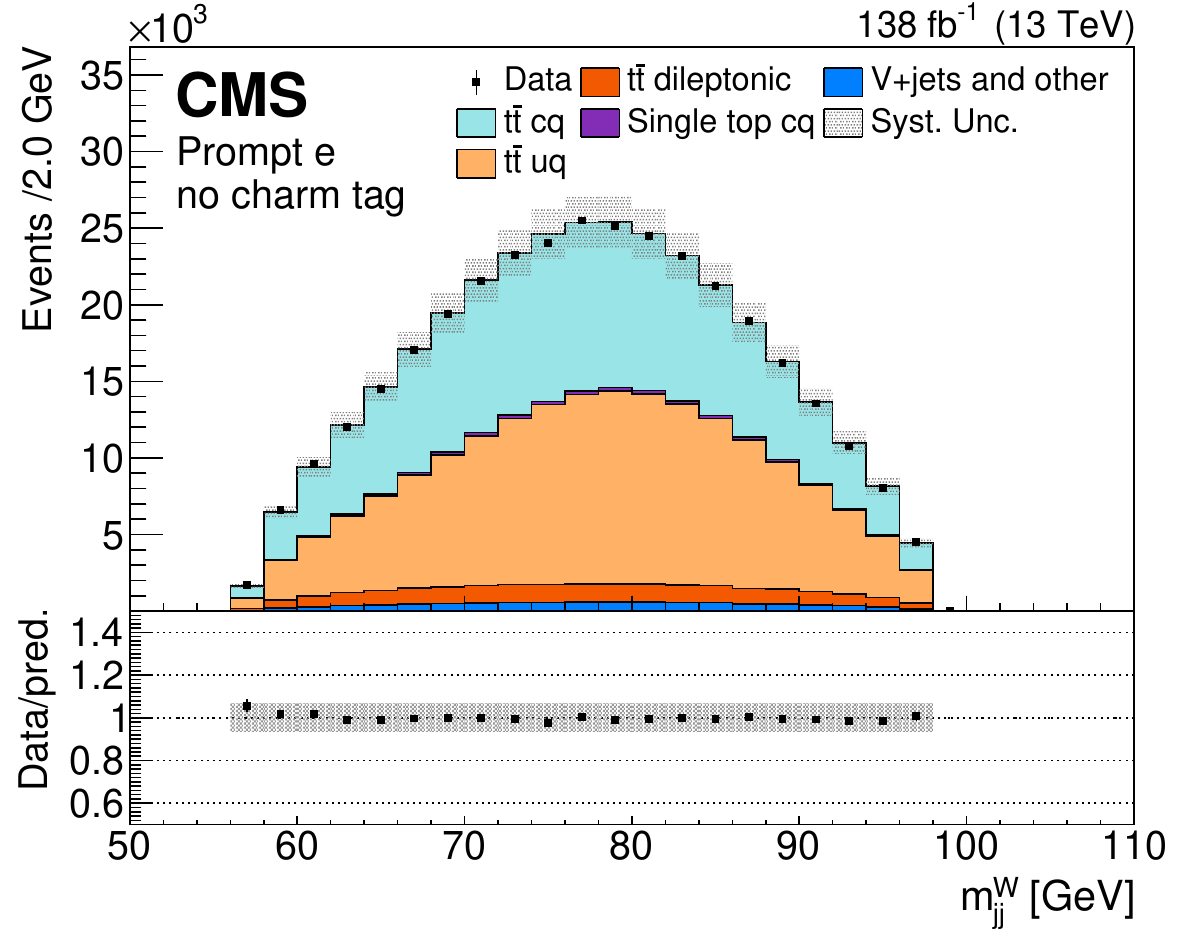}
    \includegraphics[width=0.49\textwidth]{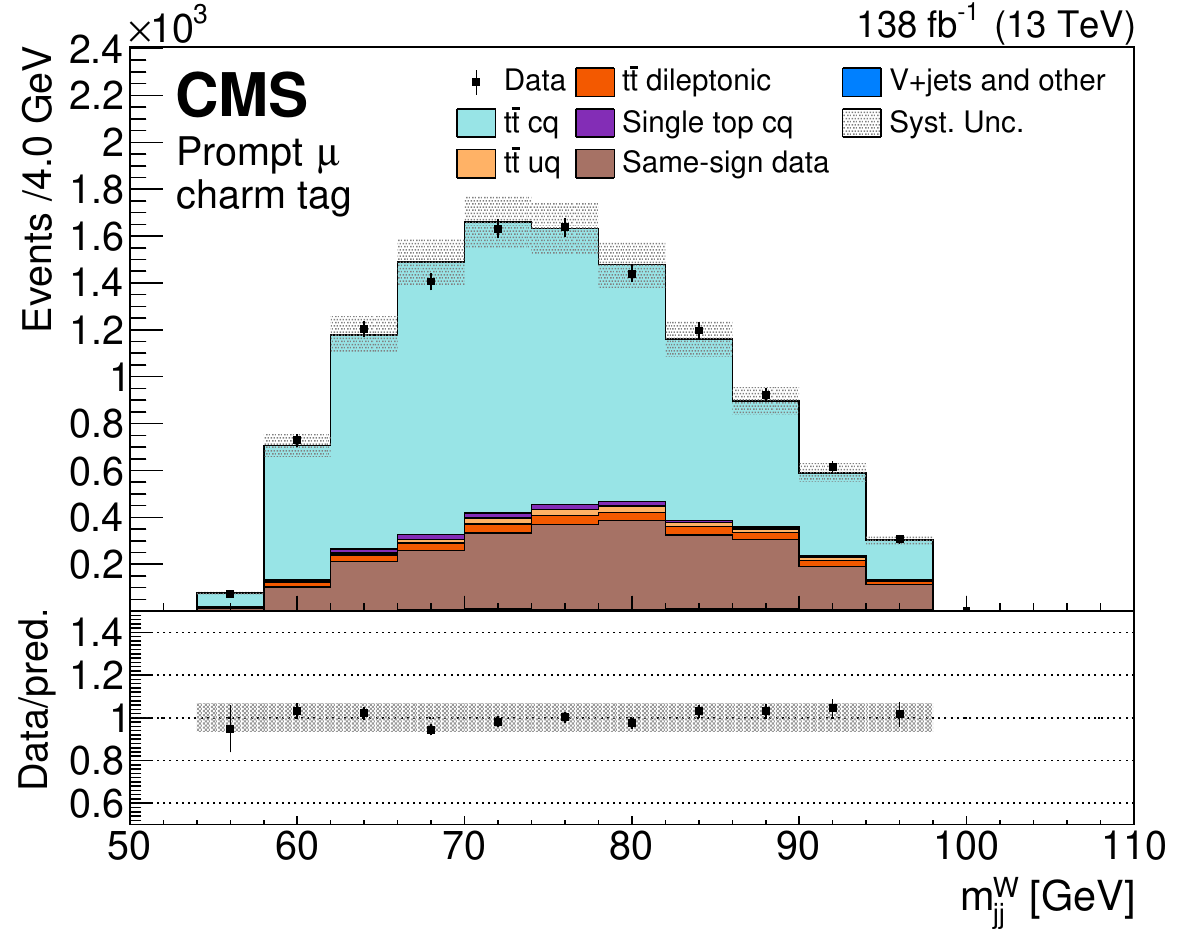}
    \includegraphics[width=0.49\textwidth]{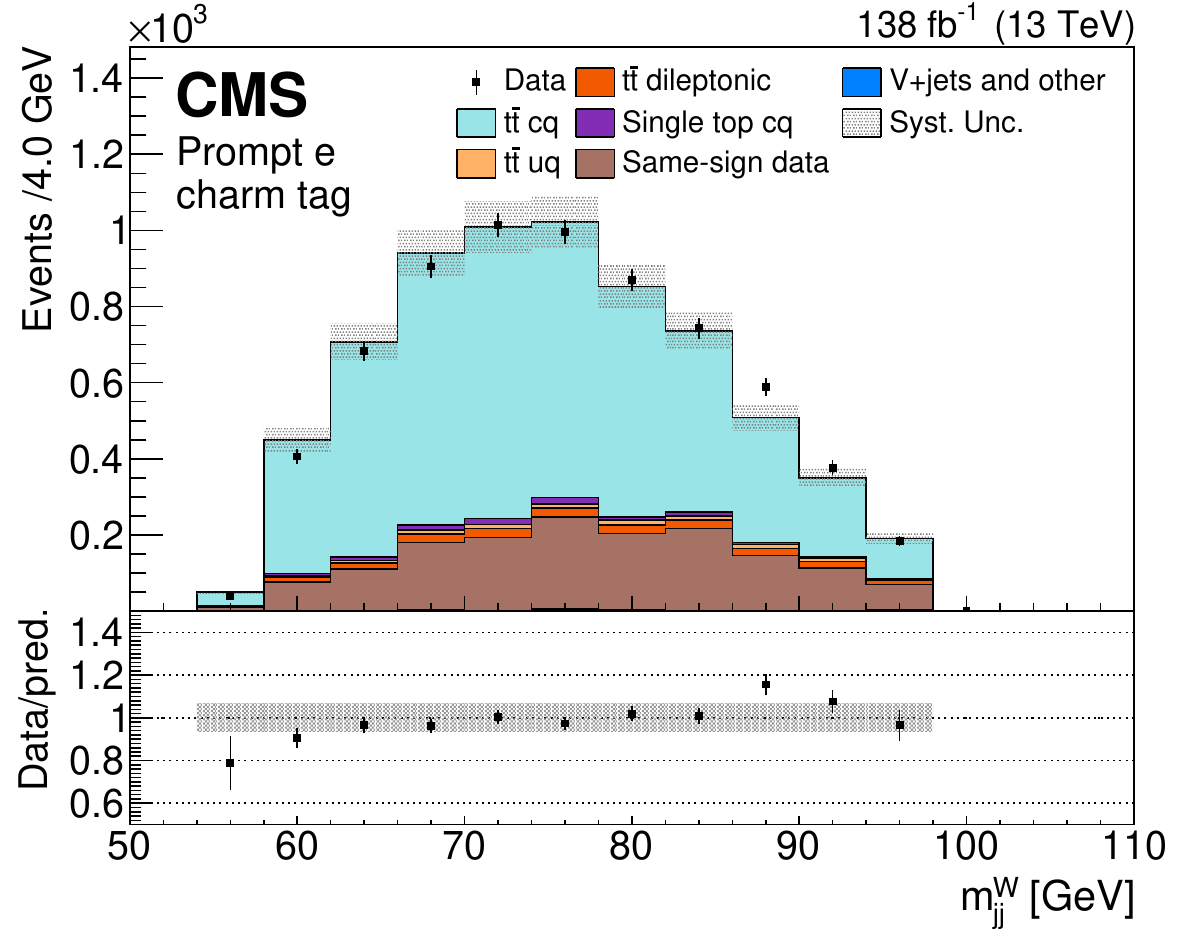} 
    \caption{Invariant mass of the two jets associated with the \PW boson, for the four event categories entering the fit: prompt muon and no charm tag (upper left), prompt electron and no charm tag (upper right), prompt muon and charm tag (lower left), prompt electron and charm tag (lower right). The measured data correspond to OS events, whereas the predictions from the simulations are \OSSS subtracted yields on top of the background prediction using the SS data. The data are displayed as points with statistical error bars. The gray band in the predictions represents the pre-fit systematic uncertainties. The ratios of data to the expected yields are given at the bottom of each panel together with the statistical and systematic uncertainties.\label{fig:mjj} }
\end{figure*}

\begin{figure*}[h!]
    \centering
    \includegraphics[width=0.49\textwidth]{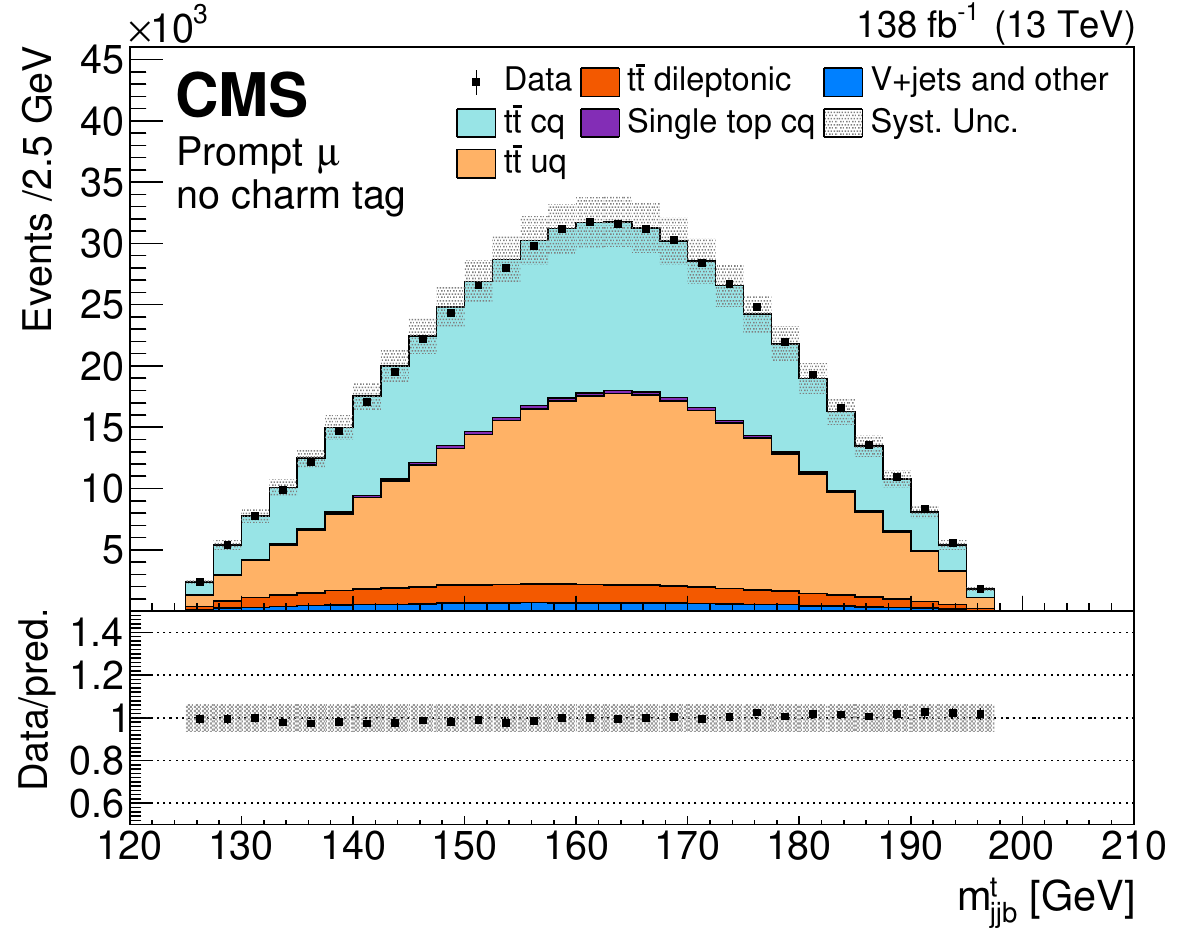}
    \includegraphics[width=0.49\textwidth]{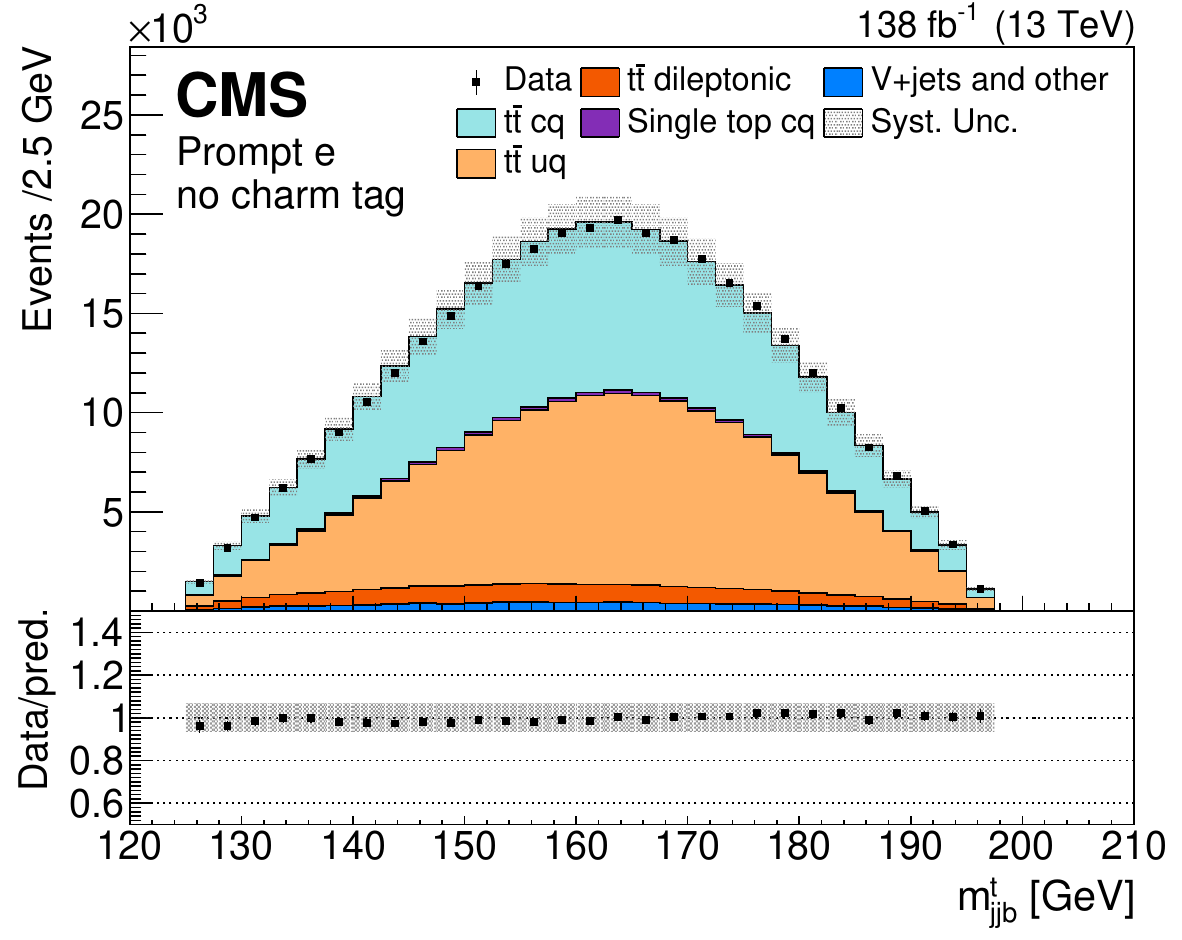}
    \includegraphics[width=0.49\textwidth]{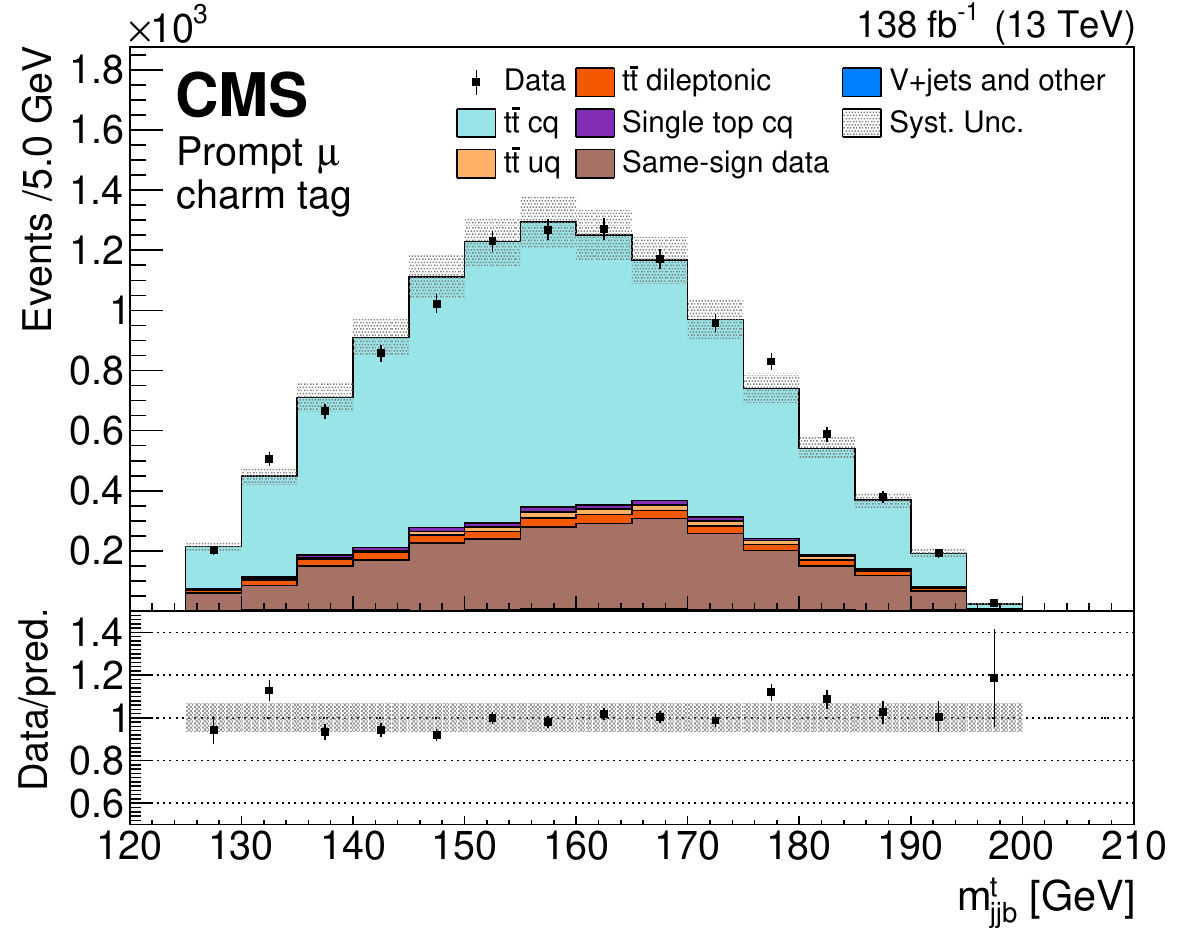}
    \includegraphics[width=0.49\textwidth]{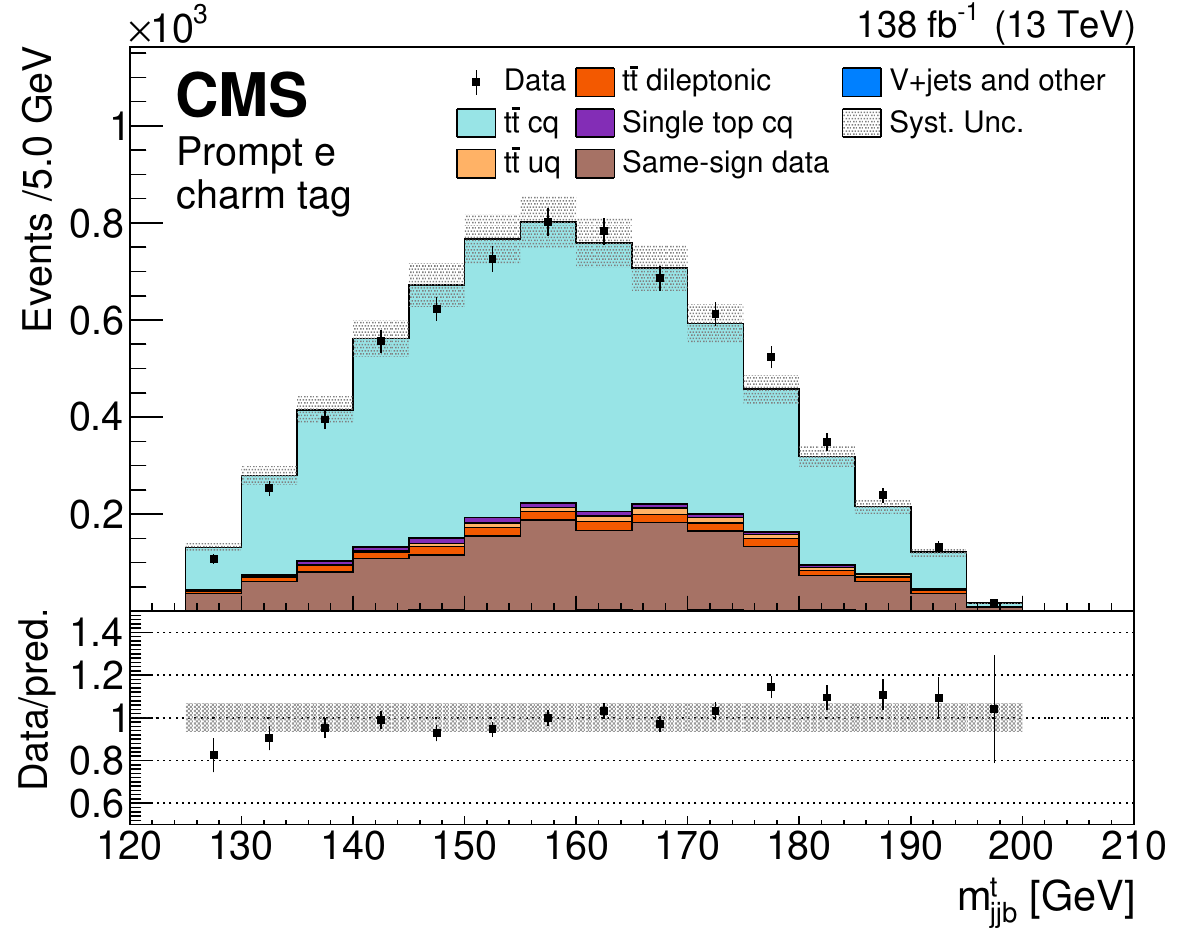} 
    \caption{Invariant mass of the three jets associated with the top quark, for the four event categories entering the fit: prompt muon and no charm tag (upper left), prompt electron and no charm tag (upper right), prompt muon and charm tag (lower left), prompt electron and charm tag (lower right). The measured data correspond to OS events, whereas the predictions from the simulations are \OSSS subtracted yields on top of the background prediction using the SS data. The data are displayed as points with statistical error bars. The gray band in the predictions represents the pre-fit systematic uncertainties. The ratios of data to the expected yields are given at the bottom of each panel together with the statistical and systematic uncertainties.\label{fig:mjjb} }
\end{figure*}

To illustrate the good description of the data by the simulation, we display in Figs.~\ref{fig:mjj} and~\ref{fig:mjjb}, for the four event categories used in the fit, the post-fit distributions of the invariant mass of the two jets associated with the \PW boson, and the invariant mass of the three jets associated with the top quark. The counting fit slightly adjusts the normalization of the simulation predictions while leaving the shapes of the distributions unchanged.

Using Eq.~(\ref{eq:CKM}) and the CMS measurement of the sum of squared elements in the first two rows of the CKM matrix ($1.984 \pm 0.021$, determined from the measured \PW boson leptonic decay BFs~\cite{CMSWBR}), the sum of squared elements in the second row of the CKM matrix can be derived. The obtained value, $\Sigma = 0.970 \pm 0.041$, provides a consistency test of the CKM matrix unitarity.  
In addition, using the measured value of $\Sigma$ and the world-average values of $\abs{V_{\PQc\PQd}} = 0.221 \pm 0.004$ and $\abs{V_{\PQc\PQb}} = 0.0408 \pm 0.0014$~\cite{PDG24}, a value of $\abs{V_{\PQc\PQs}}=0.959 \pm 0.021$ is obtained.  
This value is less precise than the current world-average value, $\abs{V_{\PQc\PQs}}= 0.975\pm 0.006$~\cite{PDG24}, determined from leptonic and semileptonic charm hadron decays, but it provides an independent measurement from hadronic \PW boson decays. 

\section{Summary~\label{sec:summ}}
The most precise measurement to date of the \PW boson hadronic decay branching fraction ratio \Rcw = \brratio is reported. The measurement is based on a data sample collected by the CMS experiment at a center-of-mass energy of 13\TeV in 2016--2018 at the CERN LHC with an integrated luminosity of 138\fbinv.  The large cross section of top quark-antiquark production at the LHC offers a sizable high-purity sample of \PW bosons suitable for this measurement. Events with one prompt charged lepton (electron or muon) and at least four jets, two of them tagged as bottom quark jets, are analyzed. Charm jets are tagged using the presence of a muon inside the jet. This charm tagging method enables the selection of a sample of charm jets with a low level of background that is precisely determined from data. The measured \Rcw value is $0.489 \pm 0.020$, in good agreement with the standard model prediction. The relative precision of the measurement of 4\%,  limited by the systematic uncertainty in the charm tagging efficiency, is improved by a factor of two compared with the current world-average value.   

From the \Rcw measurement, the sum of squared elements in the second row of the Cabibbo--Kobayashi--Maskawa (CKM) matrix, $0.970 \pm 0.041$, and the CKM matrix element $\abs{V_{\PQc\PQs}}=0.959 \pm 0.021$ are derived. These results provide a consistency test of the CKM unitarity and a measurement of $\abs{V_{\PQc\PQs}}$ from hadronic \PW boson decays. 

\begin{acknowledgments}
  We congratulate our colleagues in the CERN accelerator departments for the excellent performance of the LHC and thank the technical and administrative staffs at CERN and at other CMS institutes for their contributions to the success of the CMS effort. In addition, we gratefully acknowledge the computing centers and personnel of the Worldwide LHC Computing Grid and other centers for delivering so effectively the computing infrastructure essential to our analyses. Finally, we acknowledge the enduring support for the construction and operation of the LHC, the CMS detector, and the supporting computing infrastructure provided by the following funding agencies: SC (Armenia), BMBWF and FWF (Austria); FNRS and FWO (Belgium); CNPq, CAPES, FAPERJ, FAPERGS, and FAPESP (Brazil); MES and BNSF (Bulgaria); CERN; CAS, MoST, and NSFC (China); MINCIENCIAS (Colombia); MSES and CSF (Croatia); RIF (Cyprus); SENESCYT (Ecuador); ERC PRG, RVTT3 and MoER TK202 (Estonia); Academy of Finland, MEC, and HIP (Finland); CEA and CNRS/IN2P3 (France); SRNSF (Georgia); BMBF, DFG, and HGF (Germany); GSRI (Greece); NKFIH (Hungary); DAE and DST (India); IPM (Iran); SFI (Ireland); INFN (Italy); MSIP and NRF (Republic of Korea); MES (Latvia); LMTLT (Lithuania); MOE and UM (Malaysia); BUAP, CINVESTAV, CONACYT, LNS, SEP, and UASLP-FAI (Mexico); MOS (Montenegro); MBIE (New Zealand); PAEC (Pakistan); MES and NSC (Poland); FCT (Portugal);  MESTD (Serbia); MCIN/AEI and PCTI (Spain); MOSTR (Sri Lanka); Swiss Funding Agencies (Switzerland); MST (Taipei); MHESI and NSTDA (Thailand); TUBITAK and TENMAK (Turkey); NASU (Ukraine); STFC (United Kingdom); DOE and NSF (USA).
  
  \hyphenation{Rachada-pisek} Individuals have received support from the Marie-Curie program and the European Research Council and Horizon 2020 Grant, contract Nos.\ 675440, 724704, 752730, 758316, 765710, 824093, 101115353, 101002207, and COST Action CA16108 (European Union); the Leventis Foundation; the Alfred P.\ Sloan Foundation; the Alexander von Humboldt Foundation; the Science Committee, project no. 22rl-037 (Armenia); the Belgian Federal Science Policy Office; the Fonds pour la Formation \`a la Recherche dans l'Industrie et dans l'Agriculture (FRIA-Belgium); the F.R.S.-FNRS and FWO (Belgium) under the ``Excellence of Science -- EOS" -- be.h project n.\ 30820817; the Beijing Municipal Science \& Technology Commission, No. Z191100007219010 and Fundamental Research Funds for the Central Universities (China); the Ministry of Education, Youth and Sports (MEYS) of the Czech Republic; the Shota Rustaveli National Science Foundation, grant FR-22-985 (Georgia); the Deutsche Forschungsgemeinschaft (DFG), among others, under Germany's Excellence Strategy -- EXC 2121 ``Quantum Universe" -- 390833306, and under project number 400140256 - GRK2497; the Hellenic Foundation for Research and Innovation (HFRI), Project Number 2288 (Greece); the Hungarian Academy of Sciences, the New National Excellence Program - \'UNKP, the NKFIH research grants K 131991, K 133046, K 138136, K 143460, K 143477, K 146913, K 146914, K 147048, 2020-2.2.1-ED-2021-00181, TKP2021-NKTA-64, and 2021-4.1.2-NEMZ\_KI-2024-00036 (Hungary); the Council of Science and Industrial Research, India; ICSC -- National Research Center for High Performance Computing, Big Data and Quantum Computing and FAIR -- Future Artificial Intelligence Research, funded by the NextGenerationEU program (Italy); the Latvian Council of Science; the Ministry of Education and Science, project no. 2022/WK/14, and the National Science Center, contracts Opus 2021/41/B/ST2/01369 and 2021/43/B/ST2/01552 (Poland); the Funda\c{c}\~ao para a Ci\^encia e a Tecnologia, grant CEECIND/01334/2018 (Portugal); the National Priorities Research Program by Qatar National Research Fund; MCIN/AEI/10.13039/501100011033, ERDF ``a way of making Europe", and the Programa Estatal de Fomento de la Investigaci{\'o}n Cient{\'i}fica y T{\'e}cnica de Excelencia Mar\'{\i}a de Maeztu, grant MDM-2017-0765 and Programa Severo Ochoa del Principado de Asturias (Spain); the Chulalongkorn Academic into Its 2nd Century Project Advancement Project, and the National Science, Research and Innovation Fund via the Program Management Unit for Human Resources \& Institutional Development, Research and Innovation, grant B39G670016 (Thailand); the Kavli Foundation; the Nvidia Corporation; the SuperMicro Corporation; the Welch Foundation, contract C-1845; and the Weston Havens Foundation (USA).
\end{acknowledgments}
\bibliography{auto_generated}
\cleardoublepage \appendix\section{The CMS Collaboration \label{app:collab}}\begin{sloppypar}\hyphenpenalty=5000\widowpenalty=500\clubpenalty=5000\input{SMP-24-009-public-authorlist.tex}\end{sloppypar}
\end{document}

%% file: SMP-24-009-public-authorlist.tex
\cmsinstitute{Yerevan Physics Institute, Yerevan, Armenia}
{\tolerance=6000
A.~Hayrapetyan, A.~Tumasyan\cmsAuthorMark{1}\cmsorcid{0009-0000-0684-6742}
\par}
\cmsinstitute{Institut f\"{u}r Hochenergiephysik, Vienna, Austria}
{\tolerance=6000
W.~Adam\cmsorcid{0000-0001-9099-4341}, J.W.~Andrejkovic, L.~Benato\cmsorcid{0000-0001-5135-7489}, T.~Bergauer\cmsorcid{0000-0002-5786-0293}, S.~Chatterjee\cmsorcid{0000-0003-2660-0349}, K.~Damanakis\cmsorcid{0000-0001-5389-2872}, M.~Dragicevic\cmsorcid{0000-0003-1967-6783}, P.S.~Hussain\cmsorcid{0000-0002-4825-5278}, M.~Jeitler\cmsAuthorMark{2}\cmsorcid{0000-0002-5141-9560}, N.~Krammer\cmsorcid{0000-0002-0548-0985}, A.~Li\cmsorcid{0000-0002-4547-116X}, D.~Liko\cmsorcid{0000-0002-3380-473X}, I.~Mikulec\cmsorcid{0000-0003-0385-2746}, J.~Schieck\cmsAuthorMark{2}\cmsorcid{0000-0002-1058-8093}, R.~Sch\"{o}fbeck\cmsAuthorMark{2}\cmsorcid{0000-0002-2332-8784}, D.~Schwarz\cmsorcid{0000-0002-3821-7331}, M.~Sonawane\cmsorcid{0000-0003-0510-7010}, W.~Waltenberger\cmsorcid{0000-0002-6215-7228}, C.-E.~Wulz\cmsAuthorMark{2}\cmsorcid{0000-0001-9226-5812}
% ...
\par}
\vskip\cmsinstskip
\dag:~Deceased\\
$^{1}$Also at Yerevan State University, Yerevan, Armenia\\
$^{2}$Also at TU Wien, Vienna, Austria\\
% ...


%% file: SMP-24-009_temp.bbl
\providecommand{\href}[2]{#2}\begingroup\raggedright\begin{thebibliography}{10}%
\makeatletter
\providecommand{\hrefCMSnoop }[0]{\@secondoftwo}%
\makeatother
\providecommand{\doi}{\texttt{doi:}\begingroup \urlstyle{tt}\Url}

\bibitem{Cabibbo:1963}
\hrefCMSnoop {}{N.~Cabibbo, ``Unitary symmetry and leptonic decays'',} \textit{
  Phys. Rev. Lett.} \textbf{ 10} (1963) 531,
  \href{http://dx.doi.org/10.1103/PhysRevLett.10.531}{\doi{10.1103/PhysRevLett.10.531}}.

\bibitem{Kobayashi:1973}
\hrefCMSnoop {}{M.~Kobayashi and T.~Maskawa, ``{CP} violation in the
  renormalizable theory of weak interaction'',} \textit{ Prog. Theor. Phys.}
  \textbf{ 49} (1973) 652,
  \href{http://dx.doi.org/10.1143/PTP.49.652}{\doi{10.1143/PTP.49.652}}.

\bibitem{PDG24}
\hrefCMSnoop {}{{Particle Data Group}, S.~Navas { et~al.}, ``Review of particle
  physics'',} \textit{ Phys. Rev. D} \textbf{ 110} (2024) 030001,
  \href{http://dx.doi.org/10.1103/PhysRevD.110.030001}{\doi{10.1103/PhysRevD.110.030001}}.

\bibitem{RcwALEPH}
\hrefCMSnoop {}{{ALEPH} Collaboration, ``A direct measurement of
  $\abs{V_{\PQc\PQs}}$ in hadronic {W} decays using a charm tag'',} \textit{
  Phys. Lett. B} \textbf{ 465} (1999) 349,
  \href{http://dx.doi.org/10.1016/S0370-2693(99)01088-6}{\doi{10.1016/S0370-2693(99)01088-6}}.

\bibitem{RcwOPAL}
\hrefCMSnoop {}{{OPAL} Collaboration, ``A measurement of the rate of charm
  production in {W} decays'',} \textit{ Phys. Lett. B} \textbf{ 490} (2000) 71,
  \href{http://dx.doi.org/10.1016/S0370-2693(00)00971-0}{\doi{10.1016/S0370-2693(00)00971-0}},
  \href{http://www.arXiv.org/abs/hep-ex/0009020}{\texttt{arXiv:hep-ex/0009020}}.

\bibitem{EPJC78}
\hrefCMSnoop {}{{CMS Collaboration}, ``{Measurement of associated {Z} + charm
  production in proton-proton collisions at $\ensuremath{\sqrt{s}} =
  8$~TeV}'',} \textit{ Eur. Phys. J. C} \textbf{ 78} (2018) 287,
  \href{http://dx.doi.org/10.1140/epjc/s10052-018-5752-x}{\doi{10.1140/epjc/s10052-018-5752-x}},
  \href{http://www.arXiv.org/abs/1711.02143}{\texttt{arXiv:1711.02143}}.

\bibitem{CMS-PAPER-SMP-12-002}
\hrefCMSnoop {}{{CMS Collaboration}, ``{Measurement of associated {W} + charm
  production in pp collisions at $\ensuremath{\sqrt{s}} = 7$ TeV}'',} \textit{
  JHEP} \textbf{ 02} (2014) 013,
  \href{http://dx.doi.org/10.1007/JHEP02(2014)013}{\doi{10.1007/JHEP02(2014)013}},
\href{http://www.arXiv.org/abs/1310.1138}{\texttt{arXiv:1310.1138}}.

\bibitem{CMS-PAPER-SMP-18-013}
\hrefCMSnoop {}{{CMS Collaboration}, ``{Measurements of the production of a {W}
  boson in association with a charm quark in proton-proton collisions at
  $\sqrt{s}=8$~TeV}'',} \textit{ Eur. Phys. J. C} \textbf{ 82} (2022) 1094,
  \href{http://dx.doi.org/10.1140/epjc/s10052-022-10897-7}{\doi{10.1140/epjc/s10052-022-10897-7}},
  \href{http://www.arXiv.org/abs/2112.00895}{\texttt{arXiv:2112.00895}}.

\bibitem{CMS-PAPER-SMP-21-005}
\hrefCMSnoop {}{{CMS Collaboration}, ``{Measurements of the production cross
  section for a {W} boson in association with a charm quark in proton-proton
  collisions at $\sqrt{s}=13$~TeV}'',} \textit{ Eur. Phys. J. C} \textbf{ 84}
  (2024) 27,
  \href{http://dx.doi.org/10.1140/epjc/s10052-023-12258-4}{\doi{10.1140/epjc/s10052-023-12258-4}},
  \href{http://www.arXiv.org/abs/2308.02285v2}{\texttt{arXiv:2308.02285v2}}.

\bibitem{CMS:2008xjf}
\hrefCMSnoop {}{{CMS Collaboration}, ``The {CMS} experiment at the {CERN}
  {LHC}'',} \textit{ JINST} \textbf{ 3} (2008) S08004,
  \href{http://dx.doi.org/10.1088/1748-0221/3/08/S08004}{\doi{10.1088/1748-0221/3/08/S08004}}.

\bibitem{CMS:2020cmk}
\hrefCMSnoop {}{{CMS Collaboration}, ``{Performance of the CMS Level-1 trigger
  in proton-proton collisions at $\sqrt{s} = 13$\,TeV}'',} \textit{ JINST}
  \textbf{ 15} (2020) P10017,
  \href{http://dx.doi.org/10.1088/1748-0221/15/10/P10017}{\doi{10.1088/1748-0221/15/10/P10017}},
  \href{http://www.arXiv.org/abs/2006.10165}{\texttt{arXiv:2006.10165}}.

\bibitem{CMS:2016ngn}
\hrefCMSnoop {}{{CMS Collaboration}, ``{The CMS trigger system}'',} \textit{
  JINST} \textbf{ 12} (2017) P01020,
  \href{http://dx.doi.org/10.1088/1748-0221/12/01/P01020}{\doi{10.1088/1748-0221/12/01/P01020}},
\href{http://www.arXiv.org/abs/1609.02366}{\texttt{arXiv:1609.02366}}.

\bibitem{CMS:2017yfk}
\hrefCMSnoop {}{{CMS Collaboration}, ``Particle-flow reconstruction and global
  event description with the {CMS} detector'',} \textit{ JINST} \textbf{ 12}
  (2017) P10003,
  \href{http://dx.doi.org/10.1088/1748-0221/12/10/P10003}{\doi{10.1088/1748-0221/12/10/P10003}},
\href{http://www.arXiv.org/abs/1706.04965}{\texttt{arXiv:1706.04965}}.

\bibitem{CMS:2020uim}
\hrefCMSnoop {}{{CMS Collaboration}, ``Electron and photon reconstruction and
  identification with the {CMS} experiment at the {CERN} {LHC}'',} \textit{
  JINST} \textbf{ 16} (2021) P05014,
  \href{http://dx.doi.org/10.1088/1748-0221/16/05/P05014}{\doi{10.1088/1748-0221/16/05/P05014}},
  \href{http://www.arXiv.org/abs/2012.06888}{\texttt{arXiv:2012.06888}}.

\bibitem{CMS:2018rym}
\hrefCMSnoop {}{{CMS Collaboration}, ``{Performance of the CMS muon detector
  and muon reconstruction with proton-proton collisions at $\sqrt{s}=$ 13
  TeV}'',} \textit{ JINST} \textbf{ 13} (2018) P06015,
  \href{http://dx.doi.org/10.1088/1748-0221/13/06/P06015}{\doi{10.1088/1748-0221/13/06/P06015}},
  \href{http://www.arXiv.org/abs/1804.04528}{\texttt{arXiv:1804.04528}}.

\bibitem{CMS-TDR-15-02}
\href {http://cds.cern.ch/record/2020886}{{CMS Collaboration}, ``Technical
  proposal for the {phase-II} upgrade of the {Compact Muon Solenoid}''.} CMS
  Technical Proposal, 2015.
\newblock CERN-LHCC-2015-010, CMS-TDR-15-02. \url
  {http://cds.cern.ch/record/2020886}.

\bibitem{Cacciari:2008gp}
\hrefCMSnoop {}{M.~Cacciari, G.~P. Salam, and G.~Soyez, ``{The anti-\kt jet
  clustering algorithm}'',} \textit{ JHEP} \textbf{ 04} (2008) 063,
  \href{http://dx.doi.org/10.1088/1126-6708/2008/04/063}{\doi{10.1088/1126-6708/2008/04/063}},
  \href{http://www.arXiv.org/abs/0802.1189}{\texttt{arXiv:0802.1189}}.

\bibitem{Cacciari:2011ma}
\hrefCMSnoop {}{M.~Cacciari, G.~P. Salam, and G.~Soyez, ``{FastJet user
  manual}'',} \textit{ Eur. Phys. J. C} \textbf{ 72} (2012) 1896,
  \href{http://dx.doi.org/10.1140/epjc/s10052-012-1896-2}{\doi{10.1140/epjc/s10052-012-1896-2}},
\href{http://www.arXiv.org/abs/1111.6097}{\texttt{arXiv:1111.6097}}.

\bibitem{CMS-PU}
\hrefCMSnoop {}{{CMS Collaboration}, ``Pileup mitigation at {CMS} in {$13\TeV$}
  data'',} \textit{ JINST} \textbf{ 15} (2020) P09018,
  \href{http://dx.doi.org/10.1088/1748-0221/15/09/p09018}{\doi{10.1088/1748-0221/15/09/p09018}},
  \href{http://www.arXiv.org/abs/2003.00503}{\texttt{arXiv:2003.00503}}.

\bibitem{CMS:2016lmd}
\hrefCMSnoop {}{{CMS Collaboration}, ``{Jet energy scale and resolution in the
  CMS experiment in pp collisions at 8 TeV}'',} \textit{ JINST} \textbf{ 12}
  (2017) P02014,
  \href{http://dx.doi.org/10.1088/1748-0221/12/02/P02014}{\doi{10.1088/1748-0221/12/02/P02014}},
\href{http://www.arXiv.org/abs/1607.03663}{\texttt{arXiv:1607.03663}}.

\bibitem{CMS:2019ctu}
\hrefCMSnoop {}{{CMS Collaboration}, ``Performance of missing transverse
  momentum reconstruction in proton-proton collisions at $\sqrt{s} = 13$\,{TeV}
  using the {CMS} detector'',} \textit{ JINST} \textbf{ 14} (2019) P07004,
  \href{http://dx.doi.org/10.1088/1748-0221/14/07/P07004}{\doi{10.1088/1748-0221/14/07/P07004}},
\href{http://www.arXiv.org/abs/1903.06078}{\texttt{arXiv:1903.06078}}.

\bibitem{CMS-PAPER-EWK-10-005}
\hrefCMSnoop {}{{CMS Collaboration}, ``Measurement of the inclusive {W} and {Z}
  production cross sections in pp collisions at $\sqrt{s}=7$ {TeV}'',} \textit{
  JHEP} \textbf{ 10} (2011) 132,
  \href{http://dx.doi.org/10.1007/JHEP10(2011)132}{\doi{10.1007/JHEP10(2011)132}},
\href{http://www.arXiv.org/abs/1107.4789}{\texttt{arXiv:1107.4789}}.

\bibitem{Nason:2004rx}
\hrefCMSnoop {}{P.~Nason, ``{A new method for combining NLO QCD with shower
  Monte Carlo algorithms}'',} \textit{ JHEP} \textbf{ 11} (2004) 40,
  \href{http://dx.doi.org/10.1088/1126-6708/2004/11/040}{\doi{10.1088/1126-6708/2004/11/040}},
\href{http://www.arXiv.org/abs/hep-ph/0409146}{\texttt{arXiv:hep-ph/0409146}}.

\bibitem{Frixione:2007vw}
\hrefCMSnoop {}{S.~Frixione, P.~Nason, and C.~Oleari, ``{Matching NLO QCD
  computations with parton shower simulations: the POWHEG method}'',} \textit{
  JHEP} \textbf{ 11} (2007) 70,
  \href{http://dx.doi.org/10.1088/1126-6708/2007/11/070}{\doi{10.1088/1126-6708/2007/11/070}},
\href{http://www.arXiv.org/abs/0709.2092}{\texttt{arXiv:0709.2092}}.

\bibitem{Alioli:2010xd}
\hrefCMSnoop {}{S.~Alioli, P.~Nason, C.~Oleari, and E.~Re, ``{A general
  framework for implementing NLO calculations in shower Monte Carlo programs:
  the POWHEG BOX}'',} \textit{ JHEP} \textbf{ 06} (2010) 043,
  \href{http://dx.doi.org/10.1007/JHEP06(2010)043}{\doi{10.1007/JHEP06(2010)043}},
\href{http://www.arXiv.org/abs/1002.2581}{\texttt{arXiv:1002.2581}}.

\bibitem{Alwall}
J.~Alwall\hrefCMSnoop {}{ { et~al.}, ``The automated computation of tree-level
  and next-to-leading order differential cross sections, and their matching to
  parton shower simulations'',} \textit{ JHEP} \textbf{ 07} (2014) 079,
  \href{http://dx.doi.org/10.1007/JHEP07(2014)079}{\doi{10.1007/JHEP07(2014)079}},
  \href{http://www.arXiv.org/abs/1405.0301}{\texttt{arXiv:1405.0301}}.

\bibitem{Sjostrand:2014zea}
T.~Sj{\"o}strand\hrefCMSnoop {}{ { et~al.}, ``{An Introduction to PYTHIA
  8.2}'',} \textit{ Comput. Phys. Commun.} \textbf{ 191} (2015) 159,
  \href{http://dx.doi.org/10.1016/j.cpc.2015.01.024}{\doi{10.1016/j.cpc.2015.01.024}},
\href{http://www.arXiv.org/abs/1410.3012}{\texttt{arXiv:1410.3012}}.

\bibitem{NNPDF31nlo}
\hrefCMSnoop {}{{NNPDF} Collaboration, ``{Parton distributions from
  high-precision collider data}'',} \textit{ Eur. Phys. J. C} \textbf{ 77}
  (2017) 663,
  \href{http://dx.doi.org/10.1140/epjc/s10052-017-5199-5}{\doi{10.1140/epjc/s10052-017-5199-5}},
  \href{http://www.arXiv.org/abs/1706.00428}{\texttt{arXiv:1706.00428}}.

\bibitem{CP5}
\hrefCMSnoop {}{{CMS Collaboration}, ``{Extraction and validation of a new set
  of CMS {\sc Pythia8} tunes from underlying-event measurements}'',} \textit{
  Eur. Phys. J. C} \textbf{ 80} (2020) 4,
  \href{http://dx.doi.org/10.1140/epjc/s10052-019-7499-4}{\doi{10.1140/epjc/s10052-019-7499-4}},
  \href{http://www.arXiv.org/abs/1903.12179}{\texttt{arXiv:1903.12179}}.

\bibitem{Agostinelli:2002hh}
\hrefCMSnoop {}{{GEANT4} Collaboration, ``{\GEANTfour}--a simulation
  toolkit'',} \textit{ Nucl. Instrum. Meth. A} \textbf{ 506} (2003) 250,
\href{http://dx.doi.org/10.1016/S0168-9002(03)01368-8}{\doi{10.1016/S0168-9002(03)01368-8}}.

\bibitem{CMS-PAPER-BTV-16-002}
\hrefCMSnoop {}{{CMS Collaboration}, ``{Identification of heavy-flavour jets
  with the CMS detector in pp collisions at 13 TeV}'',} \textit{ JINST}
  \textbf{ 13} (2018) P05011,
  \href{http://dx.doi.org/10.1088/1748-0221/13/05/P05011}{\doi{10.1088/1748-0221/13/05/P05011}},
  \href{http://www.arXiv.org/abs/1712.07158}{\texttt{arXiv:1712.07158}}.

\bibitem{Bols:2020bkb}
E.~Bols\hrefCMSnoop {}{ { et~al.}, ``Jet flavour classification using
  {DeepJet}'',} \textit{ JINST} \textbf{ 15} (2020) P12012,
  \href{http://dx.doi.org/10.1088/1748-0221/15/12/P12012}{\doi{10.1088/1748-0221/15/12/P12012}},
  \href{http://www.arXiv.org/abs/2008.10519}{\texttt{arXiv:2008.10519}}.

\bibitem{Lisovyi:2015uqa}
\hrefCMSnoop {}{M.~Lisovyi, A.~Verbytskyi, and O.~Zenaiev, ``{Combined analysis
  of charm-quark fragmentation-fraction measurements}'',} \textit{ Eur. Phys.
  J. C} \textbf{ 76} (2016) 397,
  \href{http://dx.doi.org/10.1140/epjc/s10052-016-4246-y}{\doi{10.1140/epjc/s10052-016-4246-y}},
\href{http://www.arXiv.org/abs/1509.01061}{\texttt{arXiv:1509.01061}}.

\bibitem{CMS-DP-2018-058}
\href {http://cds.cern.ch/record/2646773}{{CMS Collaboration}, ``{Performance
  of the DeepJet b tagging algorithm using 41.9/fb of data from proton-proton
  collisions at 13 TeV with Phase 1 CMS detector}'',} CMS Detector Performance
  Summary CMS-DP-2018-058, 2018.

\bibitem{CMS:2018mlc}
\hrefCMSnoop {}{{CMS Collaboration}, ``{Measurement of the inelastic
  proton-proton cross section at $\sqrt{s}=13$ {TeV}}'',} \textit{ JHEP}
  \textbf{ 07} (2018) 161,
  \href{http://dx.doi.org/10.1007/JHEP07(2018)161}{\doi{10.1007/JHEP07(2018)161}},
  \href{http://www.arXiv.org/abs/1802.02613}{\texttt{arXiv:1802.02613}}.

\bibitem{CMS-LUM-17-003}
\hrefCMSnoop {}{{CMS Collaboration}, ``Precision luminosity measurement in
  proton-proton collisions at $\sqrt{s} =$ 13 {TeV} in 2015 and 2016 at
  {CMS}'',} \textit{ Eur. Phys. J. C} \textbf{ 81} (2021) 800,
  \href{http://dx.doi.org/10.1140/epjc/s10052-021-09538-2}{\doi{10.1140/epjc/s10052-021-09538-2}},
  \href{http://www.arXiv.org/abs/2104.01927}{\texttt{arXiv:2104.01927}}.

\bibitem{CMS-PAS-LUM-17-004}
\href {https://cds.cern.ch/record/2621960/}{{CMS Collaboration}, ``{CMS}
  luminosity measurement for the 2017 data-taking period at $\sqrt{s}$ = 13
  {TeV}''.} CMS Physics Analysis Summary, 2018.
\newblock CMS-PAS-LUM-17-004. \url {https://cds.cern.ch/record/2621960/}.

\bibitem{CMS-PAS-LUM-18-002}
\href {https://cds.cern.ch/record/2676164/}{{CMS Collaboration}, ``{CMS}
  luminosity measurement for the 2018 data-taking period at $\sqrt{s}$ = 13
  {TeV}''.} CMS Physics Analysis Summary, 2019.
\newblock CMS-PAS-LUM-18-002. \url {https://cds.cern.ch/record/2676164/}.

\bibitem{Czakon:2011xx}
\hrefCMSnoop {}{M.~Czakon and A.~Mitov, ``{Top++: A program for the calculation
  of the top-pair cross-section at hadron colliders}'',} \textit{ Comput. Phys.
  Commun.} \textbf{ 185} (2014) 2930,
  \href{http://dx.doi.org/10.1016/j.cpc.2014.06.021}{\doi{10.1016/j.cpc.2014.06.021}},
\href{http://www.arXiv.org/abs/1112.5675}{\texttt{arXiv:1112.5675}}.

\bibitem{Campbell:2020fhf}
\hrefCMSnoop {}{J.~Campbell, T.~Neumann, and Z.~Sullivan, ``{Single-top-quark
  production in the $t$-channel at NNLO}'',} \textit{ JHEP} \textbf{ 02} (2021)
  040,
  \href{http://dx.doi.org/10.1007/JHEP02(2021)040}{\doi{10.1007/JHEP02(2021)040}},
  \href{http://www.arXiv.org/abs/2012.01574}{\texttt{arXiv:2012.01574}}.

\bibitem{Kidonakis:2021vob}
\hrefCMSnoop {}{N.~Kidonakis and N.~Yamanaka, ``{Higher-order corrections for
  $tW$ production at high-energy hadron colliders}'',} \textit{ JHEP} \textbf{
  05} (2021) 278,
  \href{http://dx.doi.org/10.1007/JHEP05(2021)278}{\doi{10.1007/JHEP05(2021)278}},
  \href{http://www.arXiv.org/abs/2102.11300}{\texttt{arXiv:2102.11300}}.

\bibitem{CMS-W-Z}
\hrefCMSnoop {}{{CMS Collaboration}, ``{Measurement of the inclusive cross
  sections for W and Z boson production in proton-proton collisions at
  $\sqrt{s}$ = 5.02 and 13 TeV}'',} 2024.
  \href{http://www.arXiv.org/abs/2408.03744}{\texttt{arXiv:2408.03744}}.
  Submitted to \textit{JHEP}.

\bibitem{CMS-PAPER-SMP-18-004}
\hrefCMSnoop {}{{CMS Collaboration}, ``${W}^{+}{W}^{\ensuremath{-}}$ boson pair
  production in proton-proton collisions at $\sqrt{s}=13\text{ }\text{
  }\mathrm{TeV}$'',} \textit{ Phys. Rev. D} \textbf{ 102} (2020) 092001,
  \href{http://dx.doi.org/10.1103/PhysRevD.102.092001}{\doi{10.1103/PhysRevD.102.092001}},
  \href{http://www.arXiv.org/abs/2009.00119}{\texttt{arXiv:2009.00119}}.

\bibitem{CMS-PAPER-SMP-19-001}
\hrefCMSnoop {}{{CMS Collaboration}, ``{Measurements of ${\mathrm{p}}
  {\mathrm{p}} \rightarrow {\mathrm{Z}} {\mathrm{Z}}$ production cross sections
  and constraints on anomalous triple gauge couplings at $\sqrt{s} = 13\,\text
  {TeV} $}'',} \textit{ Eur. Phys. J. C} \textbf{ 81} (2021) 200,
  \href{http://dx.doi.org/10.1140/epjc/s10052-020-08817-8}{\doi{10.1140/epjc/s10052-020-08817-8}},
  \href{http://www.arXiv.org/abs/2009.01186}{\texttt{arXiv:2009.01186}}.

\bibitem{CMS-PAPER-SMP-20-014}
\hrefCMSnoop {}{{CMS Collaboration}, ``{Measurement of the inclusive and
  differential WZ production cross sections, polarization angles, and triple
  gauge couplings in pp collisions at $\sqrt{s}$ = 13 TeV}'',} \textit{ JHEP}
  \textbf{ 07} (2022) 32,
  \href{http://dx.doi.org/10.1007/JHEP07(2022)032}{\doi{10.1007/JHEP07(2022)032}},
  \href{http://www.arXiv.org/abs/2110.11231}{\texttt{arXiv:2110.11231}}.

\bibitem{PhysRevD.95.092001}
\hrefCMSnoop {}{{CMS Collaboration} Collaboration, ``Measurement of
  differential cross sections for top quark pair production using the
  $\text{lepton}+\text{jets}$ final state in proton-proton collisions at 13
  tev'',} \textit{ Phys. Rev. D} \textbf{ 95} (May, 2017) 092001,
  \href{http://dx.doi.org/10.1103/PhysRevD.95.092001}{\doi{10.1103/PhysRevD.95.092001}},
  \href{http://www.arXiv.org/abs/1610.04191v2}{\texttt{arXiv:1610.04191v2}}.

\bibitem{charmDELPHI}
\hrefCMSnoop {}{{DELPHI} Collaboration, ``{A measurement of {D} meson
  production in ${Z}^{0}$ hadronic decays}'',} \textit{ Z. Phys. C} \textbf{
  59} (1993) 533,
  \href{http://dx.doi.org/10.1007/BF01562545}{\doi{10.1007/BF01562545}}.

\bibitem{charmOPAL}
\hrefCMSnoop {}{{OPAL} Collaboration, ``{A measurement of the production of
  ${D}^{*\pm}$ mesons on the ${Z}^{0}$ resonance}'',} \textit{ Z. Phys. C}
  \textbf{ 67} (1995) 27,
  \href{http://dx.doi.org/10.1007/BF01564819}{\doi{10.1007/BF01564819}}.

\bibitem{charmALEPH}
\hrefCMSnoop {}{{ALEPH} Collaboration, ``{Study of charm production in {Z}
  decays}'',} \textit{ Eur. Phys. J. C} \textbf{ 16} (2000) 597,
  \href{http://dx.doi.org/10.1007/s100520000421}{\doi{10.1007/s100520000421}},
  \href{http://www.arXiv.org/abs/hep-ex/9909032}{\texttt{arXiv:hep-ex/9909032}}.

\bibitem{combine}
\hrefCMSnoop {}{{CMS Collaboration}, ``The {CMS} statistical analysis and
  combination tool: {\textsc{Combine}}'',} \textit{ Comput. Softw. Big Sci.}
  \textbf{ 8} (2024) 19,
  \href{http://dx.doi.org/10.1007/s41781-024-00121-4}{\doi{10.1007/s41781-024-00121-4}},
  \href{http://www.arXiv.org/abs/2404.06614}{\texttt{arXiv:2404.06614}}.

\bibitem{CMSWBR}
\hrefCMSnoop {}{{CMS Collaboration}, ``Precision measurement of the {W} boson
  decay branching fractions in proton-proton collisions at $\sqrt{s}=13\text{
  }\text{ }\mathrm{TeV}$'',} \textit{ Phys. Rev. D} \textbf{ 105} (2022)
  072008,
  \href{http://dx.doi.org/10.1103/PhysRevD.105.072008}{\doi{10.1103/PhysRevD.105.072008}},
  \href{http://www.arXiv.org/abs/2201.07861}{\texttt{arXiv:2201.07861}}.

\end{thebibliography}\endgroup
